\documentclass{emulateapj}
\usepackage[sort&compress]{natbib}
\usepackage{aas_macros}
\usepackage{bm}
\usepackage{mathtools}

\usepackage{hyperref}
\begin{document} 

\title{An adaptive scheduling tool to optimize measurements to reach a scientific objective: methodology and application to the measurements of stellar orbits in the Galactic Center}

\author{A. Hees\altaffilmark{1,2}, A. Dehghanfar\altaffilmark{2,3}, T. Do\altaffilmark{2}, A. M. Ghez\altaffilmark{2},  G. D. Martinez\altaffilmark{2}, R. Campbell\altaffilmark{4},  J. R. Lu\altaffilmark{5}}

\affil{{$^1$SYRTE, Observatoire de Paris, Universit\'e PSL, CNRS, Sorbonne Universit\'e, LNE, 61 avenue de l'Observatoire, 75014 Paris, France} \\
{$^2$Department of Physics and Astronomy, University of California, Los Angeles, Los Angeles, CA 90095, USA } \\
{$^3$Institut de Plan\'etologie et d'Astrophysique de Grenoble (IPAG), 120 rue de la Piscine, 38041 Grenoble, France } \\
$^4${W. M. Keck Observatory, 65-1120 Mamalahoa Highway, Kamuela, HI 96743, USA} \\
$^5${Astronomy Department, University of California, Berkeley, CA 94720, USA}}

\begin{abstract}
\noindent In various fields of physics and astronomy, access to experimental facilities or to telescopes is becoming more and more competitive and limited. It becomes therefore important to optimize the type of measurements and their scheduling to reach a given scientific objective and to increase the chances of success of a scientific project. In this communication, extending the work of \citet{ford:2008aa} and of~\citet{loredo:2012aa}, we present an efficient adaptive scheduling tool aimed at prioritzing measurements in order to reach a scientific goal. The algorithm, based on the Fisher matrix, can be applied to a wide class of measurements. We present this algorithm in detail and discuss some practicalities such as systematic errors or measurements losses due to contigencies (such as weather, experimental failure, \dots). As an illustration, we consider measurements of the short-period star S0-2 in our Galactic Center. We show that the radial velocity measurements at the two turning points of the radial velocity curve are more powerful for detecting the gravitational redshift than measurements at the maximal relativistic signal. We also explicitly present the methodology that was used to plan measurements in order to detect the relativistic redshift considering systematics and possible measurements losses. For the future, we identify the astrometric turning points to be highly sensitive to the relativistic advance of the periastron. Finally, we also identify measurements particularly sensitive to the distance to our Galactic Center: the radial velocities around periastron and the astrometric measurements just before closest approach and at the maximal right ascension astrometric turning point.

\noindent\textit{Subject headings}: methods: data analysis -- telescopes

\end{abstract}


\section{Introduction}\label{sec:intro}
In all fields of physics, measurements and observations are costly and access to experimental facilities or telescopes becomes more and more competitive. It is therefore of prime importance to optimize measurements in order to make the most out of the time allocated to measurements or observations. Many examples of measurements that would benefit from an optimization can be found like e.g. measurements of stellar orbits in the Galactic Center (GC) in order to infer astrophysical and fundamental physics properties \citep[see e.g.][]{boehle:2016wu,gillessen:2017aa,gravity:2018aa,do:2018aa}, measurements of exoplanets in order to infer their orbital parameters \citep[see e.g.][]{konopacky:2016aa}, measurements of asteroids in order to infer orbital parameters, detect relativistic effects or infer their internal properties \citep[see e.g.][]{farnocchia:2013aa,hees:2015rc,greenberg:2017aa,verma:2017aa}, Lunar Laser Ranging measurements to infer the properties of the Moon and to perform tests of fundamental physics \citep[see e.g.][]{nordtvedt:1998aa,bourgoin:2016yu,bourgoin:2017aa,viswanathan:2018aa}, radioscience tracking of spacecraft through the Deep Space Network antennas to follow the spacecraft motion or to infer planetary properties \citep[see e.g.][]{moyer:2000uq}, specific measurements using atomic sensors in order to probe fundamental physics \citep[see e.g.][]{pihan-le-bars:2017aa,safronova:2018aa}, etc.  \ This list is not exhaustive and illustrates the wide range of projects that would benefit from an optimization of their measurements to reach a specific scientific objective.

Identifying what are the optimal measurements (type of measurements and scheduling) in order to reach a given scientific objective is not an easy task considering the large parameters space over which one needs to optimize. \citet{loredo:2003aa,loredo:2004aa,loredo:2010aa,loredo:2012aa} developed a Bayesian adaptive scheduling algorithm aiming at seeking optimized observing times for radial velocity characterization of elliptical orbits in the context of exoplanets. Another adaptive scheduling algorithm dedicated to exoplanets searches has been proposed by \citet{ford:2008aa}. Both these algorithms are based on maximizing the global information entropy increase due to one additional measurement and iterate. Maximizing the global information entropy is also known as maximum entropy sampling \citep{loredo:2012aa}. In this communication, we present another adaptive scheduling algorithm that presents three differences with respect to previous works: (i) the optimization can be done on a specific parameter of interest, for a model comparison, instead of on the global information content only ; (ii) the algorithm can be applied to any types of measurements and (iii) the algorithm is extremely fast and efficient from a computational point of view. The algorithm presented in this communication is a simple but efficient algorithm designed to prioritize measurements based on their sensitivity to a scientific objective. This algorithm relies on the computation of the Fisher matrix~\citep{fisher:1935aa} and assumes Gaussian statistics on measurements errors. Practically, this adaptive scheduling tool takes as an input existing measurements, the model that is fitted to these measurements and the criterion on which one wants to optimize and it uses a greedy algorithm to identify a set of measurements that is optimized with respect to the given criterion. It is important to emphasize that it is possible to take into account possible observational window due to seeing of a target from Earth, maintenance time of the facility, etc. In this communication, we consider three criteria for which we can prioritize our measurements: (i) the global information content inside a dataset (similarly to the work from~\citet{ford:2008aa} and \citet{loredo:2012aa}.), (ii) the uncertainty related to a scientific parameter that is estimated using the measurements or (iii) the ratio of the Bayesian evidence between two different models, a criteria used to compare two models. The set of optimized measurements will in general depend on the objective considered. The method presented in this paper is based on the computation of the Fisher matrix \citep{ly:2017aa} whose inverse can be used as an estimate of the covariance matrix (see e.g. \citet{vallisneri:2008aa,heavens:2009aa} or the Appendix from \citet{albrecht:2006aa}). Instead of computing the inverse of the Fisher matrix a large number of times, the algorithm uses an efficient technique to update the inverse of the Fisher matrix. 

The algorithm, fully described in section~\ref{sec:method}, can be applied to a wide class of measurements and of scientific projects. In section~\ref{sec:discussion}, we compare this algorithm to existing work in the literature, discuss some of its current limitations and also discuss some practicalities related to systematic effects, measurement losses due to contingencies and experimental windows. In section~\ref{sec:app}, we present an application of this tool to the case of stellar orbits measurements in the GC. It has long been anticipated that the measurements of the bright short-period star S0-2 (also named S2) orbiting the supermassive black hole (SMBH) in our GC can be used to detect and probe relativistic effects \citep{psaltis:2008uq,johannsen:2016sh,johannsen:2016uq,alexander:2005ul,jaroszynski:1998xp,rubilar:2001dq,weinberg:2005cr,zucker:2006fk,will:2008fk,angelil:2010qd,angelil:2010to,iorio:2011rc,parsa:2017aa,hees:2017aa}. The detection of such relativistic effects relies on the long time baseline measurements of this star that has been measured astrometrically since the early 1990's and spectroscopically since the early 2000's at Keck by the Galactic Center Orbit Initiative (GCOI) \citep{ghez:1998ve,ghez:2000rt,ghez:2003qv,ghez:2005dq,ghez:2005kx,ghez:2008bs,boehle:2016wu} and at the NTT and the VLT  \citep{genzel:1997zr,eckart:1997ys,schodel:2002bh,eckart:2002qf,eisenhauer:2003ty,eisenhauer:2005dz,gillessen:2009cr,gillessen:2017aa} combined with new high accurate measurements. The question of interest is: given the existing measurements of the star S0-2, what are the future measurements that will increase the chances of detection of relativistic effects? First, we show how the algorithm developed in this communication has been used to plan the 2018 GCOI observations campaign in order to successfully measure the relativistic redshift \citep{do:2018aa} (for another measurement, see also \citet{gravity:2018aa}). Secondly, we identify clearly what are the  measurements within S0-2's orbit that are relevant to detect the relativistic advance of periastron in the near future. Third, we identified what are the measurements susceptible to improve the measurement of the distance to our Galactic Center $R_0$, an important point to plan future measurements but also to search for systematic effects in existing measurements in order to understand the tension between the two estimates of $R_0$ from the Gravity collaboration \citep{gravity:2018aa} and the GCOI team \citep{do:2018aa}. Finally, our conclusions are presented in section~\ref{sec:conclusion}.

\section{Methodology}\label{sec:method}
\subsection{Formulation of the problem}\label{sec:problem}
The goal of most measurements performed is to extract some scientific information by fitting a model to the observations. We will assume the measurements errors to be normally distributed and we will note one measurement by $(m_i,t_i,\sigma_i)$ where $m_i$ is the type of measurement, $t_i$ is the time of the measurement and $\sigma_i$ is the measurement uncertainty.  In addition, we will note the model used to analyze the measurements as $M(\{ m_i,t_i \};\bm p)$ where the vector $\bm p$ denotes all the $N_\mathrm{param}$ parameters that are included in the fit. These include the parameters that are scientifically relevant but can also include other parameters such as nuisance parameters or noise parameters. In some cases, one is interested in comparing two models $M_1(\{m_i,t_i\};\bm p_1)$ and $M_2(\{m_i,t_i\};\bm p_2)$ (which do not necessarily depend on the same number of parameters) in order to identify the one favored by the data.

In addition, we will denote the set of existing measurements by $\mathcal O_\mathrm{existing}=\{(m_i,t_i,\sigma_i)\}$ and the set of possible (future) measurements by $\mathcal O_\mathrm{possible}=\{(m_j,t_j,\sigma_j)\}$. Note that the set of possible future measurements is discrete and takes into account experimental/observational constraints (observational windows, maintenance time for experimental facilities, etc\dots).

The question we are addressing in this communication can be formulated as follow:  given a set of existing measurements, how to prioritize the measurements from a set of possible future measurements in order increase the sensitivity with respect to a given scientific objective?

Three types of scientific objective will be considered: (i) a case where one is interested by maximizing the global quantity of information (the information entropy) related to the full posterior distribution, similarly to what has been developed by \citet{ford:2008aa} and \citet{loredo:2012aa}, (ii) a case where one focuses on one scientific parameter $p_k$ and where one wants to minimize the uncertainty $\delta_{p_k}$ related to the estimation of that specific parameter and (iii) a case where one wants to maximize the ratio of the Bayesian evidence between two models in order to maximize the chances to discriminate one model over the second one.

\subsection{The Fisher matrix as a tool to evaluate quickly the uncertainties and the information entropy}
The implementation of the adaptive scheduling tool presented in this paper is based on the Fisher matrix. This implicitly assumes that the posterior probability distribution function can be approximated by a multivariate normal distribution. In particular, this is the case when the errors follow a Gaussian distribution and the model can be approximated by its linearization around the maximum of likelihood. Under this assumption, the Fisher matrix is a good estimate of the inverse of the covariance matrix for unbiased estimators. The Fisher matrix, a square matrix of dimension $N_\mathrm{param}$, can be computed by (see e.g. \citet{vallisneri:2008aa,heavens:2009aa} or the appendix from \citet{albrecht:2006aa})
\begin{equation}
    \bm F = \bm P^T \cdot \bm P \, ,
\end{equation}
where $\bm P$ is the matrix of the partial derivatives of the model with respect to the parameters
\begin{equation}\label{eq:P}
    \left[\bm P \right]_{ik}=\frac{1}{\sigma_i}\frac{\partial M(i; \bm p)}{\partial p_k} \, ,
\end{equation}
where $\sigma_i$ corresponds to the uncertainty of the $i$th measurement. The matrix of the partial derivatives $\bm P$ has dimension $N_\mathrm{measurements}\times N_\mathrm{param}$.

An estimation of the uncertainties on the estimated parameters resulting from a fit of the model $M$ using a set of measurements $\{\left(m_i,t_i,\sigma_i\right)\}$ can be determined from the covariance matrix $\bm \Sigma$, which is the inverse of the Fisher matrix (see e.g. \citet{vallisneri:2008aa,heavens:2009aa} or the appendix from \citet{albrecht:2006aa})
\begin{equation}\label{eq:invFisher}
	\bm\Sigma=\bm F^{-1} \, .
\end{equation}
In particular, the marginalized uncertainties of the estimated parameters are given by the diagonal of this matrix \citep{heavens:2009aa}: 
\begin{equation}
    \delta_{p_l}=\sqrt{\Sigma_{ll}}\, .
\end{equation} 
It is important to note that this covariance matrix depends on (i) the measurements dataset (the timing and the type of measurements and their related uncertainties $\{(m_i,t_i,\sigma_i)\}$) and (ii) on the set of fitted parameters (i.e. $\bm p$).

%It can be shown that the information entropy of a multivariate distribution is directly related to the determinant of the $\bm \Sigma$ matrix through~\citep{ahmed:1989aa}
%\begin{equation}%
%	\mathcal E = -\frac{N}{2}\ln \left(2\pi e\right)-\frac{1}{2}\ln \left| \bm \Sigma  \right| \, ,
%\end{equation}
%where $N$ is the number of parameters included in the fit.

In addition, \citet{ford:2008aa} has shown that the expected difference in information entropy $\mathcal I$ that is induced by an additional  measurement $k$ is given by (see  Eq. (25) from \citet{ford:2008aa})
\begin{equation}
	\Delta_k \mathcal I = \ln \left(\frac{\sigma_\mathrm{pred;k}}{\sigma_k}\right)\, ,
\end{equation}
where $\sigma_\mathrm{pred;k}$ is the model uncertainty for the measurement $k$ obtained while not including this measurement in the fit and $\sigma_k$ is the expected measurement uncertainty. The notation $\Delta_k x$ denotes the difference in the quantity $x$ induced by adding the measurement $k$ to the existing dataset. This relationship means that the optimal measurements in order to increase the global information entropy are the ones with large model uncertainties (or, quoting \citet{loredo:2012aa}: ``we learn the most by sampling where we know the least''), which is called maximum entropy sampling \citep{sebastiani:2000aa}.

Finally, under the assumption that the posterior is a multivariate normal distribution, the model uncertainty can directly be related to the inverse of the Fisher matrix through
\begin{equation}\label{eq:sig_pred}
	\frac{\sigma_\mathrm{pred;k}^2}{\sigma_k^2} =  \tilde {\bm P}_{k}^T  \cdot \bm \Sigma \cdot \tilde {\bm P}_{k} \, ,
\end{equation}
where $\bm \Sigma$ is the covariance matrix obtained when the measurement $(m_k,t_k,\sigma_k)$ is not included in the fit and $\tilde {\bm P}_{k}$ is a column vector containing the partial derivatives of the measurement $k$ with respect to the fitted parameters $\bm p$: 
\begin{equation}\label{eq:tildeP}
	\left[\tilde {\bm P}_{k}\right]_i =\left[\bm P\right]_{ki}=\frac{1}{\sigma_k}\frac{\partial M(m_k,t_k;\bm p)}{\partial p_i} \, .
\end{equation}

One statistical tool to compare two models using a set of measurements is to use the Bayesian evidence, defined as the probability to obtain the data given a certain model (or in other words, the evidence is the normalization of the posterior) \citep[see e.g.][]{jaynes:2003aa,gregory:2010qv,gelman:2013aa,kass:1995aa,liddle:2007aa,knuth:2014aa}. Under the assumption that the prior knowledge about the two models are uniform, the Bayesian evidence is also directly proportional to the probability of a certain model given the dataset. Comparing the Bayesian evidences from fits using two different models is one way to compare these models and to discriminate them. Using the Laplace quadratic approximation, the Bayesian evidence for a model 1 is given by~\citep{liddle:2007aa}
\begin{equation}
	\mathcal E_1 = \mathcal P_\mathrm{1,max} \sqrt{\frac{\left(2\pi\right)^{N_1}}{\left|\bm \Sigma_1\right|}}\, ,
\end{equation}
where $\mathcal P_\mathrm{1,max}$ is the maximum of the posterior, $N_1$ is the number of parameters fitted in the model 1 and $\left|\bm \Sigma_1\right|$ is the determinant of the matrix $\bm \Sigma_1$. The difference of the log-evidence between two models is therefore given by
\begin{align}
	 \ln  \frac{\mathcal E_1}{\mathcal E_2} =&\ln \frac{\Pi_\mathrm{1,max}}{\Pi_\mathrm{2,max}}+\frac{N_1-N_2}{2}\ln 2\pi+ \ln \frac{\mathcal L_\mathrm{1,max}}{\mathcal L_\mathrm{2,max}} \nonumber\\
	&\quad +\frac{1}{2} \ln \left|\bm F_1\right|-\frac{1}{2}\ln \left|\bm F_2\right|\, ,
\end{align}
where $\mathcal L_\mathrm{i,max}$ and $\pi_\mathrm{i,max}$ are the likelihood and prior of the model $i$ for the parameters that maximize the posterior.
The first two terms (the ratio of the priors and the number of fitted parameters) do not depend directly on the scheduling of the measurements. On the other hand, the third part of this equation depends on the actual measurements themselves. This dependency makes it hard to find the sequence of measurements optimal to discriminate between two models (since the actual measurements are not known). One solution is to consider a Gaussian likelihood and to simulate data using one of the two models. Let us assume that model 1 is the model we are trying to confirm from the data and let us use this model to simulate the measurements so that the last equation becomes
\begin{align}
	\ln  \frac{\mathcal E_1}{\mathcal E_2}  =&\ln \frac{\Pi_\mathrm{1,max}}{\Pi_\mathrm{2,max}}+\frac{N_1-N_2}{2}\ln 2\pi+\frac{1}{2} \ln \left|\bm F_1\right|-\frac{1}{2}\ln \left|\bm F_2\right|\  \nonumber\\
	& +\sum_i \frac{\left(M_1(m_i,t_i,\bm p_1)- M_2(m_i,t_i,\bm p_2)\right)^2}{2\sigma_i^2}\, .\label{eq:model_selection}
\end{align}

In conclusion, in this section we have shown how the Fisher matrix can be used to characterize the different objectives that one may want to optimize on:
\begin{itemize}
	\item  the global information entropy: it requires to optimize over Eq.~(\ref{eq:sig_pred}) which depends on the inverse of the Fisher matrix.
	\item  the uncertainty of one specific parameter of interest ($\delta_{p_l}$): it requires to optimize over one component of the diagonal of the inverse of the Fisher matrix. 
	\item  a criterion to optimize the chances to discriminate a model $M_1$ over a model $M_2$: one suggestion is to optimize over an estimate of the difference of the log of the Bayesian evidence, given by Eq.~(\ref{eq:model_selection}). This quantity depends on the evaluation of the models themselves and on the determinant of their Fisher matrices.
\end{itemize}
In the next two subsections, we will discuss how to estimate efficiently the Fisher matrix, its determinant and its inverse.

\subsection{Computation of the partial derivatives of the model}
As can be noticed from Eq.~(\ref{eq:P}), one needs to compute the partial derivatives of the model with respect to the parameters in order to compute the Fisher matrix. Three methods can be used to compute these derivatives:
\begin{enumerate}
	\item  analytically compute  the derivatives of the model. This method has the advantage to be of quick evaluation; but it requires long calculations and code. When the model relies on the integration of a set of ordinary differential equations of motion, this method also requires  integration of the equations of variations (as an example, see section 4 from \citet{lainey:2004uq}).
	\item  use automatic differentiation. The main advantage of this method is that one does not need to derive explicitly the expression of the partial derivatives.
	\item  use finite differences. This method has the advantage of easy implementation  (especially when the model requires the integration of differential equations) but has the disadvantage that it can be numerically unstable. 
\end{enumerate}
Each of these methods has pros and cons and they need to be chosen on a case by case basis considering each specific situation.

\subsection{Update of the Fisher matrix, of its determinant and of its inverse}
One of the main step in the algorithm developed in this paper is to compute the Fisher matrix, its determinant and its inverse. Indeed, the goal of the adaptive scheduling tool is to optimize either on one diagonal component of $\bm \Sigma$ or on its determinant $\left|\bm \Sigma \right|$. Therefore, we will need to evaluate these quantities for a huge number of different combinations of measurements and it is important to compute them using a method that is computationally efficient. Recomputing the full Fisher matrix and inverting it for all the set of measurements is computationally intensive. An alternative approach is to update the the Fisher matrix as well as its determinant and its inverse to avoid to invert the full matrix again.

In other words, if we consider the Fisher matrix $\bm F^{(n)}$ obtained for a set of measurements $\mathcal O_n$, we can efficiently compute the Fisher matrix $\bm F^{(n+1)}$ which is obtained from the same set of measurements with an additional measurement $(m_k,t_k,\sigma_k)$. The procedure is straightforward and is given by
\begin{equation}
        \bm F^{(n+1)}=\bm F^{(n)}+ \tilde {\bm P}_{k}^T  \cdot \tilde {\bm P}_{k} \, .
\end{equation}

Following the same principle, the determinant of the Fisher matrix can easily be update
\begin{equation}
        \left|\bm F^{(n+1)}\right|=\left|\bm F^{(n)}\right|\left(1+\tilde {\bm P}_{k}^T  \cdot\bm \Sigma^{(n)}\cdot \tilde {\bm P}_{k}\right)\, ,
\end{equation}
where $\bm\Sigma^{(n)}=\left[\bm F^{(n)}\right]^{-1}$.

Finally, the procedure to update the covariance matrix (the inverse of the Fisher matrix) is given by
\begin{equation}\label{eq:update}
    \bm\Sigma^{(n+1)} = \bm\Sigma^{(n)} -\bm U^{(n)}_{k}\, ,
\end{equation}
where $\bm U^{(n)}_k$ is the update matrix which depends on the measurement $(m_k,t_k,\sigma_k)$ and is given by
\begin{equation}\label{eq:U}
    \bm U^{(n)}_{k}=  \frac{\left(\bm\Sigma^{(n)} . \tilde {\bm P}_{k} \right). \left(\bm\Sigma^{(n)} . \tilde {\bm P}_{k} \right)^T }{1+\tilde {\bm P}_{k}^T.\bm \Sigma^{(n)}. \tilde {\bm P}_{k}} \, ,
\end{equation}
where the vector $\tilde {\bm P}_{k}$ is given by Eq.~(\ref{eq:tildeP}).

To summarize, adding one measurement $(m_k,t_k,\sigma_k)$ to a given dataset $\mathcal O_n$ will lead to a change of the different criterion that we are optimizing on given by the following expressions:
\begin{subequations}\label{eq:eps}
\begin{itemize}
    \item optimization over the global information entropy (maximization):
    \begin{equation}
    	\varepsilon_k = \Delta_k \mathcal I = \ln \left(\tilde {\bm P}_{k}^T  \cdot \bm \Sigma^{(n)} \cdot \tilde {\bm P}_{k} \right)\, ,
    \end{equation}
    where the column vector $\tilde {\bm P}_{k}$ is given by Eq.~(\ref{eq:tildeP}).
    
    \item optimization over one specific parameter of interest (minimization of the related uncertainty):
    \begin{equation}
        \varepsilon_k = - \Delta_k \delta_{p_l}^2 =  \left[\bm U^{(n)}_k\right]_{ll} \, ,
    \end{equation}
    using Eq.~(\ref{eq:U}).
    
    \item optimization for model comparison (maximization of the ratio of the Bayesian evidence):
    \begin{align}
        \varepsilon_k = \Delta_k \ln \frac{\mathcal E_1}{\mathcal E_2}=&\frac{\left(M_1(m_k,t_k,\bm p_1)- M_2(m_k,t_k,\bm p_2)\right)^2}{2\sigma_k^2}\nonumber\\
        &+\frac{1}{2}\ln \frac{1+\tilde {\bm P}_{1,k}^T  \cdot\bm \Sigma_1^{(n)}\cdot \tilde {\bm P}_{1,k}}{1+\tilde {\bm P}_{2,k}^T  \cdot\bm \Sigma_2^{(n)}\cdot \tilde {\bm P}_{2,k}}\, ,
    \end{align}
    where the subscript 1 and 2 refers to the two models to be compared. 
\end{itemize}
\end{subequations}

\subsection{Description of the algorithm}
To start the algorithm, we need a set of existing measurements $\mathcal O_\mathrm{existing}=\left\{(m_i,t_i,\sigma_i)\right\}$ and a set of possible future measurements $\mathcal O_\mathrm{possible}=\left\{(m_j,t_j,\sigma_j)\right\}$. We also need to know what is the scientific objective for which we are optimizing for: (i) maximize the global information entropy or (ii) minimize the uncertainty related to a given parameter $p_j$ or (iii) maximize the odd ratio between two models. 

Our implementation of the adaptive scheduling tool is iterative. There are two different ways to control the number of iterations. The first way is to chose arbitrarily the desired number of measurements ($N$) and to find a scheduling for these $N$ measurements that is optimized with respect to a given criterion. The second way is to stop the iterations when we reach a given objective (for example when the uncertainty on an estimated parameter is below a given threshold). In this second case, the number of $N$ of measurements is free as well.

The algorithm consists in the following steps:
\begin{enumerate}
	\item fit the model $M(m_i,t_i;\bm p)$ to the existing measurements $\mathcal O_\mathrm{existing}$ in order to determine the optimal parameters $\bm p_\mathrm{opt}$ (in the case of models comparison, the two models need to be fitted). For this step, the actual data needs to be known and not only their uncertainties.
	\item compute the partial derivatives of the model (of the two models in case of models comparison)
	$$\frac{1}{\sigma_i} \left. \frac{\partial M(m_i,t_i;\bm p)}{\partial p_k}\right|_{\bm p_\mathrm{opt}}$$
	 for all the existing and possible future measurements. This evaluation can be computationally intensive (depending on the model) but is required only once.
	 \item compute the starting covariance matrix $\bm \Sigma^{(0)}$ from the set of existing measurements $\mathcal O_\mathrm{existing}$ by inverting the Fisher matrix, see Eq.~(\ref{eq:invFisher}).

	 \item iteratively determine the set of $N$ optimal measurements by using the following procedure (starting with $n=0$ and with an empty list of optimized measurements $\mathcal O_\mathrm{opt}=\{\}$)
	 \begin{enumerate}
	 	\item loop on all the possible future measurements $(m_k,t_k,\sigma_k)\in \mathcal O_\mathrm{possible}$ and for each of them compute the quantity $\varepsilon_k$ from Eqs.~(\ref{eq:eps}) depending on the scientific objective considered.
		\item find the measurement $(m_k,t_k,\sigma_k)$ that maximizes $\varepsilon_k$, add it to the list of optimal measurement $\mathcal O_\mathrm{opt}$ and remove it from the list of future possible measurements $\mathcal O_\mathrm{possible}$.
		\item update the covariance matrix by adding the measurement $(m_k,t_k,\sigma_k)$ using Eq.~(\ref{eq:update}), increase $n$ and iterate by going back to step (a) as long as the number of measurements in the optimal set of future observations is smaller than $N$, or as long as we have not reached the given threshold.
	 \end{enumerate}
\end{enumerate}
This algorithm is very efficient and allows one to prioritize quickly measurements based on a criteria related to a scientific objective. 

\section{Discussion}\label{sec:discussion}
The algorithm presented in the previous section can be applied to a wide class of measurements. Our algorithm presents several differences compared to existing ones (see as en example, \citet{ford:2008aa} and \citet{loredo:2012aa}). First of all, our algorithm is based on the Fisher matrix and not on a full Bayesian approach. This has the drawback that our algorithm is not guaranteed to work when posteriors are extremely non Gaussian or when they present multiple modes.  The main advantage of using the Fisher matrix relies in the efficiency and in the computational cost of the method. Indeed, we have shown in the previous section that there exist a method to efficiently update the Fisher matrix, its determinant and its inverse. This allows one to consider a huge number of possible future measurements without any problem (optimization over sets of $10^7$ possible measurements can easily be considered).

Another difference between this work and the one presented in \citet{ford:2008aa} and in \citet{loredo:2012aa} relies in the criterion that we are optimizing for. In \citet{ford:2008aa} and in \citet{loredo:2012aa}, the scheduling is performed to optimize the expected global information entropy (maximum entropy sampling). Here, we also consider two other options. The first alternative is to prioritize measurements to minimize the uncertainty of a scientific parameter. The second alternative is to prioritize measurements to maximize the chances to discriminate a model over another one. The description of the algorithm presented in the previous section includes the three options.

The algorithm presented in section~\ref{sec:method} is a greedy optimizer (similarly to the one used in section 3 from \citet{loredo:2012aa}): it chooses the measurement that leads to the best one-time increase of the metric optimized for and proceeds iteratively. While the resulting measurements sequence is optimized for a particular criterion, this algorithm is not guaranteed to converge towards the global optimal solution. Still, it has been shown in \citet{loredo:2004aa,loredo:2012aa}, using a similar optimization algorithm, the superiority of adaptive scheduling over random sampling. This superiority will also be quantified in the three examples developed in section~\ref{sec:app} below. The global optimum of the non-linear and discrete optimization problem related to the problem presented in section~\ref{sec:problem} can be determined by a brute force method whose complexity grows dramatically as $\mathcal O\left(n_p^N\right)$\footnote{This is valid when $N<<n_p$.} where $n_p$ is the number of possible future measurements and $N$ is the length of the optimal sequence. In practice, $n_p$ can be huge (as high as $10^7$), which justifies the use of a $\mathcal O(n_p)$ algorithm. In the future, it would be interesting to study more complex algorithms that could provide a better optimization compared to the greedy algorithm from section~\ref{sec:method} in a reasonable computation time complexity.
 
One input for the algorithm from section~\ref{sec:method} is the uncertainties related to possible future measurements. It is important to note that these uncertainties can vary from measurement to measurement to take into account practicalities such as, for examples, visibility from a telescope, different integration times at different observing epochs, confusion events, geometrical consideration (in the case of range measurements for example), etc. Such variations for the uncertainties are taken into account and automatically considered in the presented algorithm. On the other hand, as presented in section~\ref{sec:method}, the algorithm cannot take into account the more general case where the uncertainty for a given future measurement is given as a distribution instead of a given value. One can overcome this drawback by implementing a Monte Carlo algorithm on top of the adaptive scheduling tool. For each instance of the Monte Carlo, the uncertainties considered in the adaptive scheduling would be drawn randomly. More pragmatically and more efficiently, one can consider several cases of measurements uncertainties (e.g. a ``realistic'', an ``optimistic'' and a ``pessimistic'') and run the adaptive scheduling tool on those limited number of cases to assess the impact of the future uncertainties on the scheduling.

We would like to emphasize that practicality such as observational window, maintenance time, etc. can easily be taken into account in our approach. Indeed, our algorithm is based on a brute force exploration of all the possible future measurements. This set of possible measurements can therefore include gaps or other constraints.

It is important to note that the algorithm presented in this communication is purely based on statistical consideration. In practice, systematic effects usually play an important role when analyzing measurements and experimental losses (losses due to weather, to unexpected downtime of a telescope or of an experimental facilities, \dots) can deter the expected scientific goal. While the results that can be obtained from a scheduling tool are extremely interesting to identify optimized measurements, it is important to take into account systematics and losses when planning real measurements. When two types of measurements are considered (like the case of stellar orbits in the GC where the two types of measurements are the astrometric and the spectroscopic measurements), we suggest to plan both measurements roughly at the same epochs even though only one type of measurement is required from the adaptive scheduling tool. Using both types of measurements at the same time will help to control for systematics. 

In addition, we suggest to consider measurements losses (atmospheric losses, unexpected maintenance, etc.) by using a very simple statistical model for the losses and by running a Monte Carlo and to determine the distribution of the expected scientific results given this statistical model. This can help to estimate the number of measurements required to reach a given objective (see section~\ref{sec:red} below for a practical example).

Finally, in Eq.~(\ref{eq:P}), we implicitly assumed that the measurement error is a white noise. This can easily be generalized to the case where the measurement errors are correlated. In that case, the measurement noise is characterized by a covariance matrix $\bm C$, which is assumed to be diagonal in Eq.~(\ref{eq:P}). A covariance matrix is a definite positive matrix so that it is always possible to find a decomposition of the form
\begin{equation}
	\bm C = \bm S^T \cdot \bm S \, ,
\end{equation}
using e.g. a Cholesky decomposition. The algorithm presented in the previous section can directly be applied by replacing Eq.~(\ref{eq:P}) by
\begin{equation}
	\left[\bm P\right]_{ij} = \left[\bm S^{-1}\right]_{ik}\frac{\partial M(m_k,t_k;\bm p)}{\partial p_j} \, .
\end{equation}

\section{Application to the detection of relativistic effects on stellar orbits in the Galactic Center}\label{sec:app}
To illustrate the utility of the algorithm presented in the previous section, we use the case of the measurements of stellar orbits in the GC as an example. Of particular interest is the star S0-2 (also named S2) which has been followed astrometrically and spectroscopically for nearly 25 years \citep{eckart:1997ys,ghez:1998ve,ghez:2000rt,schodel:2002bh,eckart:2002qf,eisenhauer:2003ty,ghez:2003qv,ghez:2005dq,ghez:2005kx,eisenhauer:2005dz,ghez:2008bs,gillessen:2009cr,boehle:2016wu,gillessen:2017aa}. One of the long standing scientific objective pursued by measuring the motion of this star is to measure relativistic effects around the supermassive black hole and to test the gravitational theory of general relativity \citep{psaltis:2008uq,johannsen:2016sh,johannsen:2016uq,alexander:2005ul,jaroszynski:1998xp,rubilar:2001dq,weinberg:2005cr,zucker:2006fk,will:2008fk,angelil:2010qd,angelil:2010to,iorio:2011rc,parsa:2017aa}. Detection of relativistic effects using S0-2's measurements relies on the long standing dataset available and on new high accurate measurements. The first relativistic effects measurable using S0-2's measurements are the relativistic redshift and the relativistic advance of the periastron. Recently, the impact of the relativistic redshift on S0-2's radial velocity has been measured successfully \citep{gravity:2018aa,do:2018aa}. In this section, we will show how the adaptive scheduling tool presented in the previous section has helped the GCOI team in planning the 2018 observations campaign that led to a successful detection of the relativistic redshift \citep{do:2018aa}. In addition, we will show how the same tool can be used to identify suitable measurements in order to enhance the detection of the relativistic advance of the periastron in the near future. Finally, we will also prioritize measurements to improve the knowledge of $R_0$, our distance to the GC.

\subsection{Data and model}
The set of existing measurements used in this analysis consists in astrometric measurements performed at the Keck observatory between 1995 and 2017 and on spectroscopic measurements performed at the Keck observatory and at the Very Large Telescope (VLT) between 2000 and 2017. These measurements are reported in the literature \citep{gillessen:2017aa,do:2018aa} and here, we consider them as independent. The model $M(\{m_k,t_k\};\bm p)$ used for this analysis relies on the integration of the first post-Newtonian equations of motion and includes the R\"omer time delay and the first order relativistic redshift. This model is fully described in the Appendix of \citet{do:2018aa}. There are 15 parameters that characterize this model: the six orbital parameters for the star S0-2, the SMBH mass parameter $GM$, the distance to the GC $R_0$, the 2-D position and 3-D velocity of the SMBH, a parameter that encodes a deviation from the relativistic redshift and a parameter that encodes a deviation from the first post-Newtonian order in the equations of motion (or in other words that encodes a deviation from the relativistic advance of the periastron).

\subsection{An example of scheduling including systematics and weather losses to detect the relativistic redshift on S0-2}\label{sec:red}
In this section, we give a full example of telescope time scheduling using the tool presented in section~\ref{sec:method} including considerations related to systematics and to weather and instrumental losses. We focus on the measurement of the relativistic redshift of S0-2 and therefore fix the parameter that encodes a deviation from the first post-Newtonian equations of motion to its relativistic value (i.e. to unity). Therefore, the model contains 14 parameters. The objective is to find a measurement sequence in 2018 that would optimize the chances of a successful detection of the relativistic redshift given the set of existing measurements (up to 2017). The set of future possible measurements considered is made of two distinct sets: the first one consists in astrometric measurements for every day between March 1st 2018 and September 15th 2018 and the second one consists in radial velocity measurements for every day in the same time period. The uncertainty for the future measurements has been chosen as 20 km/s for the radial velocity and 0.9 mas for the astrometry.

\begin{figure}
\begin{center}
\includegraphics[width=.9\hsize]{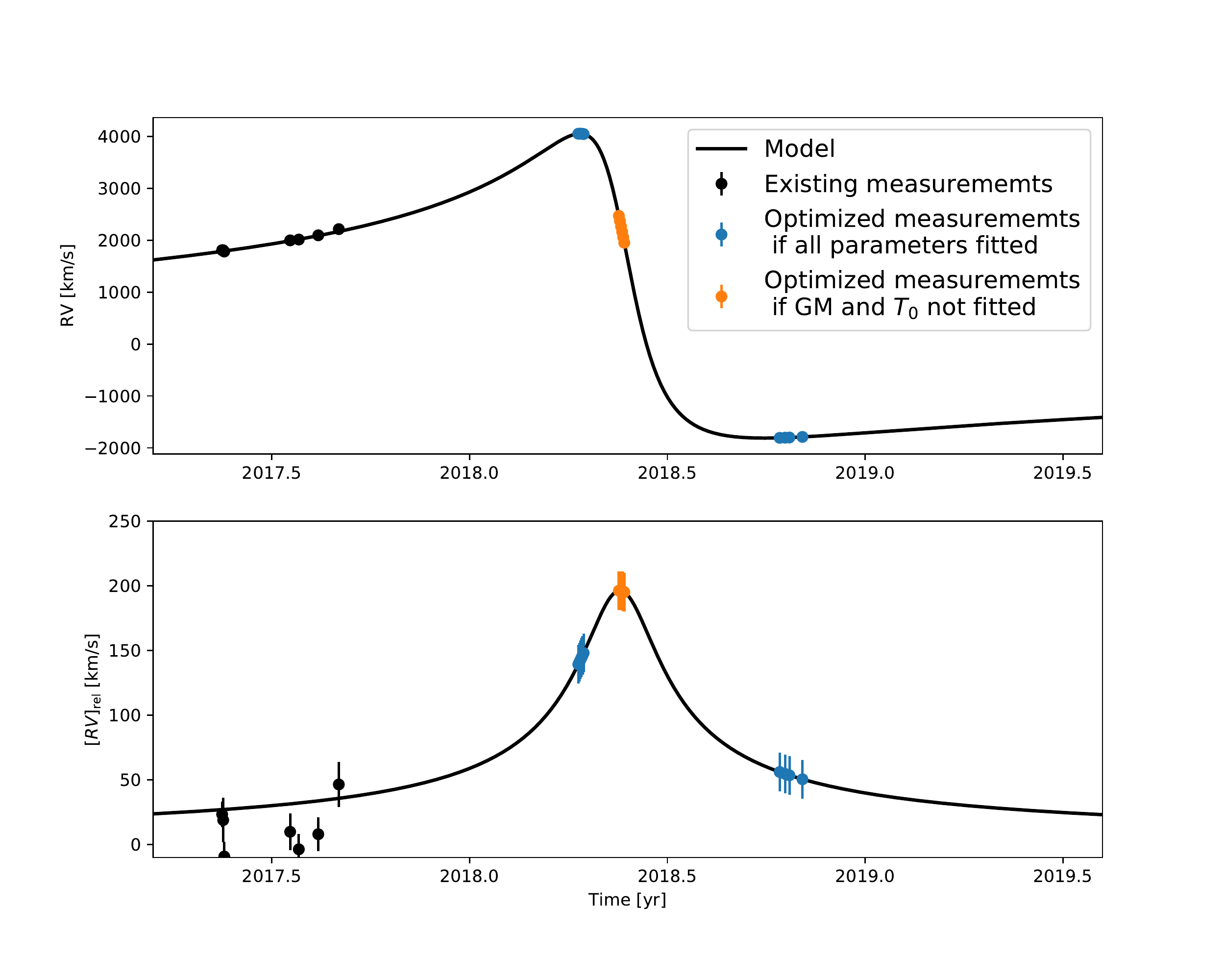}
\caption{\label{fig:cadence_redshift} Representation of the radial velocity measurements sequence in 2018 optimized to detect the relativistic redshift (see table~\ref{tab:cadence_redshift}). The top panel represents the radial velocity measurement optimized using the algorithm from section~\ref{sec:method} in order to detect the relativistic redshift. The solid curve is the radial velocity model. The bottom panel shows the relativistic contribution from the same measurements with the redshift signal. In these simulations, all measurements prior to 2018 are used (black points). The blue points show the optimized measurements sequence under the assumption that all the model parameters are fitted simultaneously. The critical radial velocity measurements are located at the turning points of the radial velocity curve and not at the maximum of the redshift because of correlations with $T_0$ and $GM$. The observations in orange represent the optimized measurements sequence obtained under the assumption that $T_0$ and $GM$ are not fitted. These measurements are located at the maximum of the redshift signal.}
\end{center}
\end{figure}

The sequence of 10 measurements optimized  in order to detect the relativistic redshift (we are optimizing on the redshift parameter) using the algorithm presented in section~\ref{sec:method} is presented in table~\ref{tab:cadence_redshift} and on figure~\ref{fig:cadence_redshift}. It is important to keep in mind that this result depends on the existing measurements: different existing measurements could have led to another sequence. This sequence also depends on the expected accuracy of the future possible measurements and on the parameters that are included in the fit.

\begin{table}[htb]
 \caption{\label{tab:cadence_redshift} Measurements sequence in 2018 optimized in order to measure the relativistic redshift given existing past measurements. Col 1: $i$ - the order of importance of the measurement. Col 2: epoch of the measurement in year. Col. 3: type of measurement. Col 4: estimation of the statistical signal to noise ratio of the relativistic redshift at the end of 2018 given the existing measurements and including $i$ new measurements. Systematics are not included here.}
	 % is used to refer this table in the text
	\centering
	\begin{tabular}{c c c c}
	\hline
	  Order of importance &  Epoch &   Type of observation  & Stat. SNR\\\hline
 1  &  2018.286  &  RV     & 4.50 \\
 2  &  2018.283  &  RV     & 4.97 \\
 3  &  2018.700  &  RV     & 5.47 \\
 4  &  2018.264  &  RV     & 6.09 \\
 5  &  2018.275  &  RV     & 6.33 \\
 6  &  2018.699  &  RV     & 6.53 \\
 7  &  2018.272  &  RV     & 6.71 \\
 8  &  2018.700  &  astro  & 6.84 \\
 9  &  2018.288  &  RV     & 6.98 \\
 10 &  2018.696  &  RV     & 7.10 \\
  \hline
 \end{tabular}
\end{table}

Two interesting facts need to be noticed about the optimized measurements sequence obtained in this analysis. First, 9 out of the 10  measurements are spectroscopic observations. This makes sense considering that the redshift impacts directly the radial velocity but not the astrometry. A more surprising fact is related to the fact that the adaptive scheduling tool does not favor measurements at the maximum of the redshift signal. Rather, it favors measurements that are at the radial velocity turning points (i.e. the maximum and minimum of the radial velocity curve, see figure~\ref{fig:cadence_redshift}). This is due to correlations between the redshift parameter and the other model parameters, in particular with $T_0$ (the time of closest approach for S0-2) and with the SMBH mass parameter $GM$. These correlations are maximal exactly when the redshift signal is maximal, making that epoch not optimal in order to measure the redshift signal. In other words, the redshift signal can easily be absorbed by a small change in $T_0$ or in $GM$. To demonstrate the effect of correlations, we run a test case and search for the optimized measurements sequence if we fit for all the model parameters except for $T_0$ and $GM$. The orange measurements from figure~\ref{fig:cadence_redshift} are the radial velocity measurements optimized for that case using the algorithm from section~\ref{sec:method}. When $T_0$ and $GM$ are not included in the fit, these measurements are all located close to the maximal of the redshift signal, as expected.

It is interesting to quantify the impact of an optimized scheduling compared to a ``naive'' one. For this, let us consider the gain in the relativistic redshift signal to noise ratio (SNR) due to one RV measurement in 2018. The redshift SNR using measurements existing up to 2017 is of 1.16. A naive scheduling consisting in performing one additional measurement at the time where the redshift signal is maximal induces a SNR increase of 4.5 \% (the SNR being increased up to 1.205), while an optimized scheduling consisting in performing the measurement at the time where the RV is maximal increases the SNR by 335 \% (the SNR value is now 4.5). This illustrates clearly the benefit of an optimized scheduling and how such a scheduling tool can help to increase the outcome of a project.

At this stage, the important point learned from our adaptive scheduling tool is that, from a statistical point of view, nearly all the power to measure the relativistic redshift comes from spectroscopic measurements around the two RV turning points. This result relies purely on statistical considerations and practicalities and systematics need to be considered as well. First of all, other scientific longer-term objectives will be pursued with S0-2's measurements. It is important to realize that measurements around closest approach, while not of prime importance for the redshift, are important to measure other scientific parameters like e.g. $R_0$ (see section~\ref{sec:R0}) and the relativistic advance of the periastron (see section~\ref{sec:GR}). Therefore, we included measurements at closest approach, an important event as well. In addition, two strategies have been used to control systematics for the relativistic redshift measurement: (i) we decided to use different telescopes (Keck and Gemini) to measure the radial velocities in order to assess hypothetical biases related to one instrument and (ii) we decided to take also astrometric measurements at the same epochs as radial velocities and vice versa (see the discussion in section~\ref{sec:discussion}). Finally, the results from Table~\ref{tab:cadence_redshift} have been obtained assuming a perfect knowledge of S0-2's orbital parameter. In practice, there is an uncertainty associated to these parameters. In particular, at the end of 2017, the uncertainty related to the 2018 time of closest approach $T_0$ was of the order of 5 days. In order to control for a possible bias in the $T_0$ estimate, we included measurements very early in 2018.

In addition, it is important to consider instrumental losses and weather losses. Using different telescopes (Keck and Gemini) for the capture of the important measurements is one way to mitigate the risk of missing important measurements because of instrumental losses (in addition to allow us to check for systematics). In addition, we increased the number of requested measurements in 2018 in order to take into account for possible measurements losses. A careful analysis of the Keck Observatory statistics shows that roughly 50 \% of GC measurements nights are lost because of weather or instrumental problems. This has been included in our scheduling by making a large number ($10^5$) of simulations where each measurement has a probability of 50\% to be lost and by assessing the distribution of the relativistic redshift signal to noise ratio at the end of 2018.

\begin{figure} 
\begin{center}
\includegraphics[width=.49\hsize]{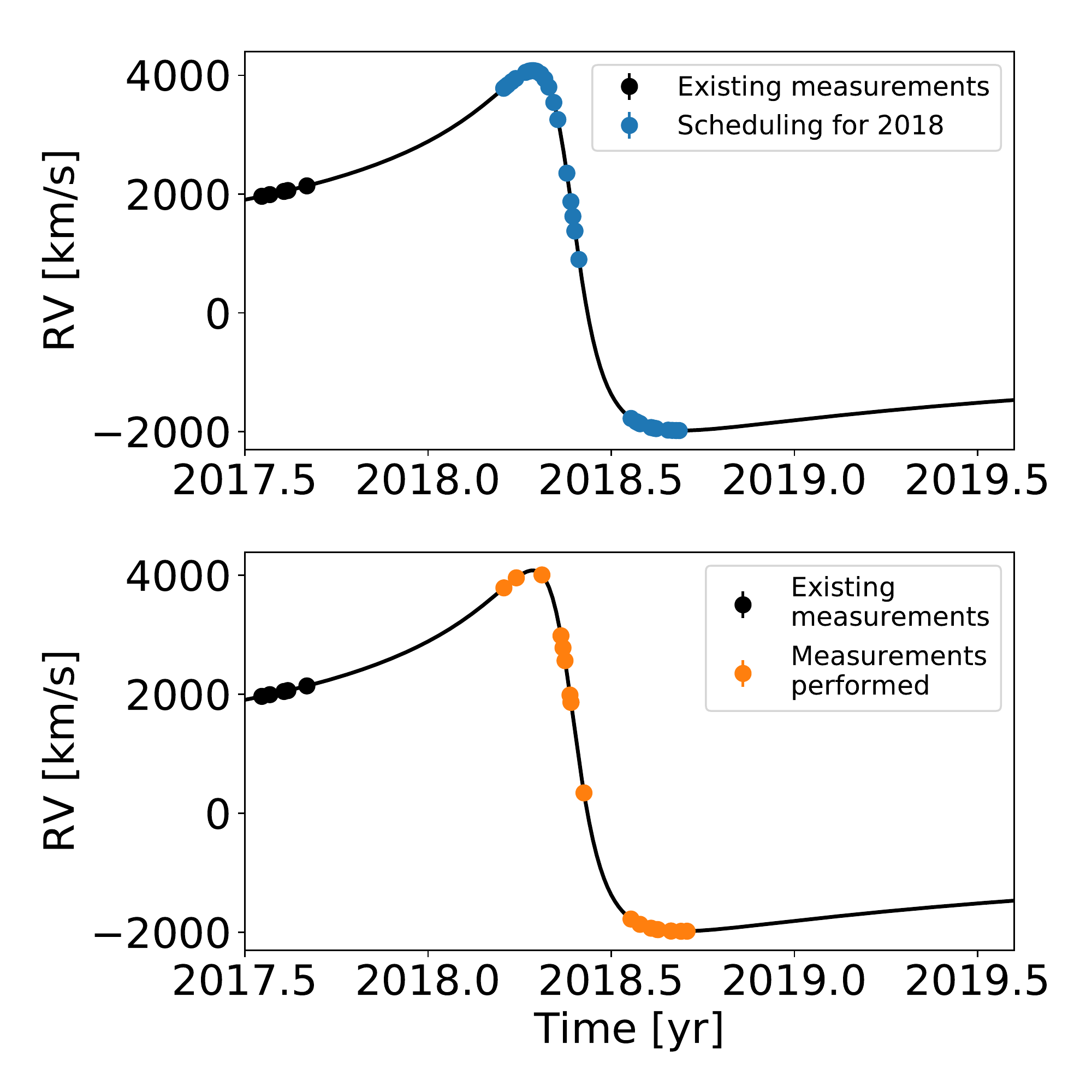} 
\includegraphics[width=.49\hsize]{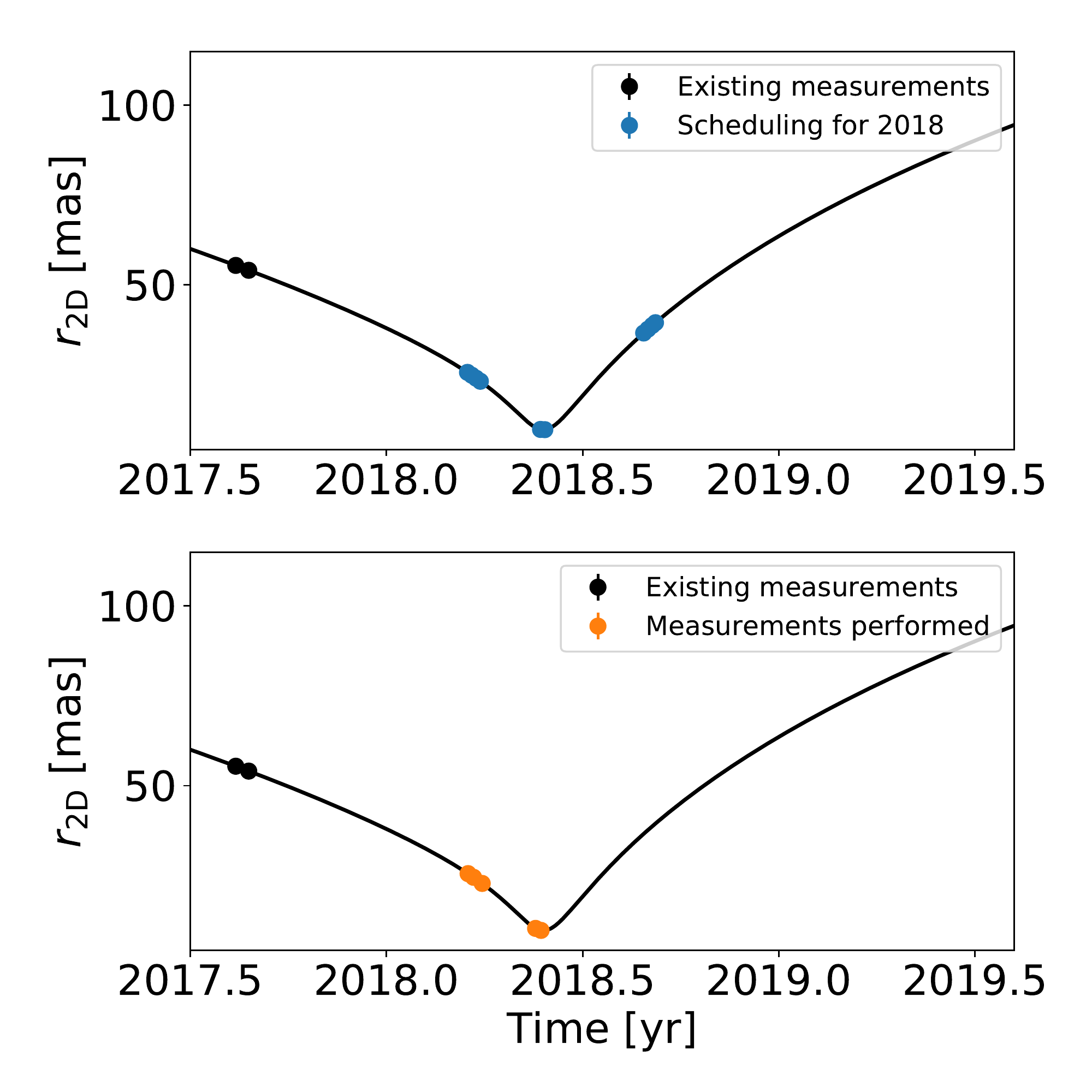}
\caption{\label{fig:redshift_scheduling} Top: representation of the scheduling for S0-2's measurements in 2018 as planned by the GCOI considering the analysis presented in this section. Bottom: actual measurements performed in 2018, see \cite{do:2018aa}. Left: radial velocity (sepctroscopic measurements). Right: astrometric measurements presented as a 2-D radial projection between S0-2 and the center of the reference frame \citet{sakai:2019aa}.}
\end{center}
\end{figure}

\begin{figure}
\begin{center}
\includegraphics[width=.37\hsize]{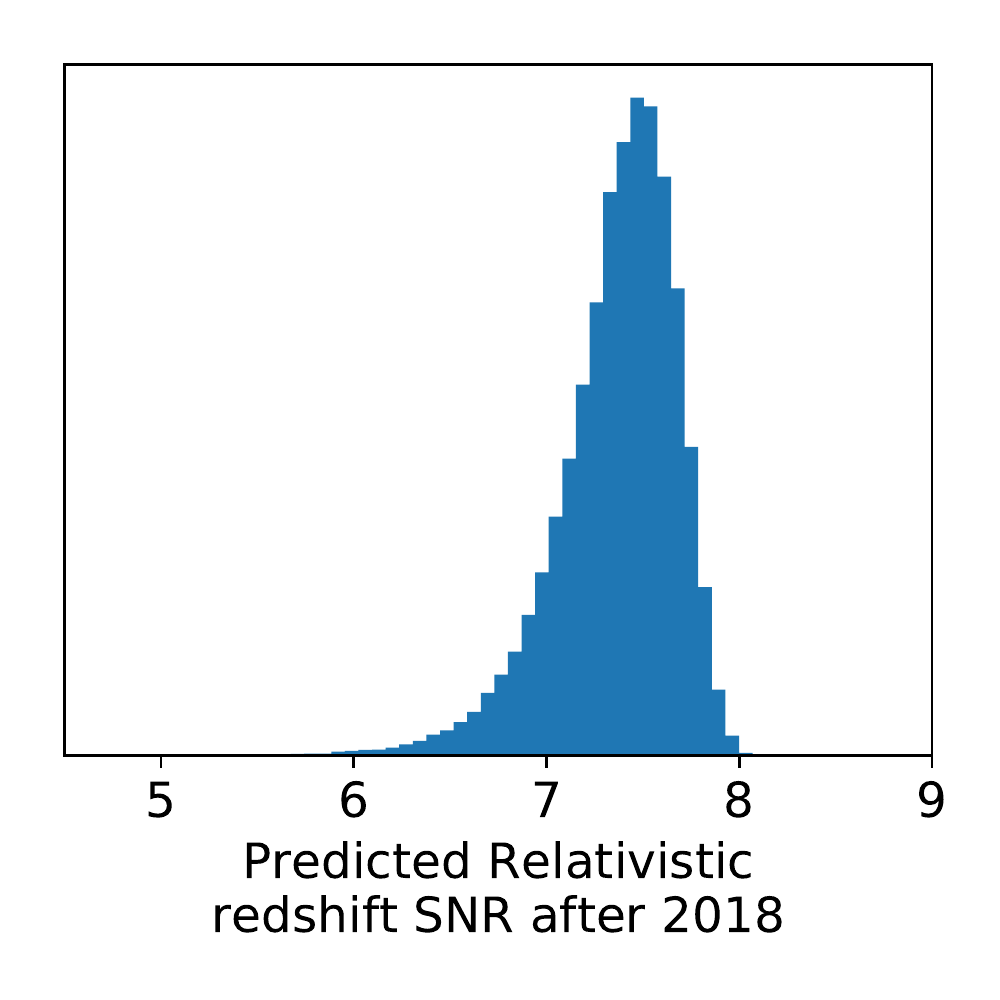}
\includegraphics[width=.52\hsize]{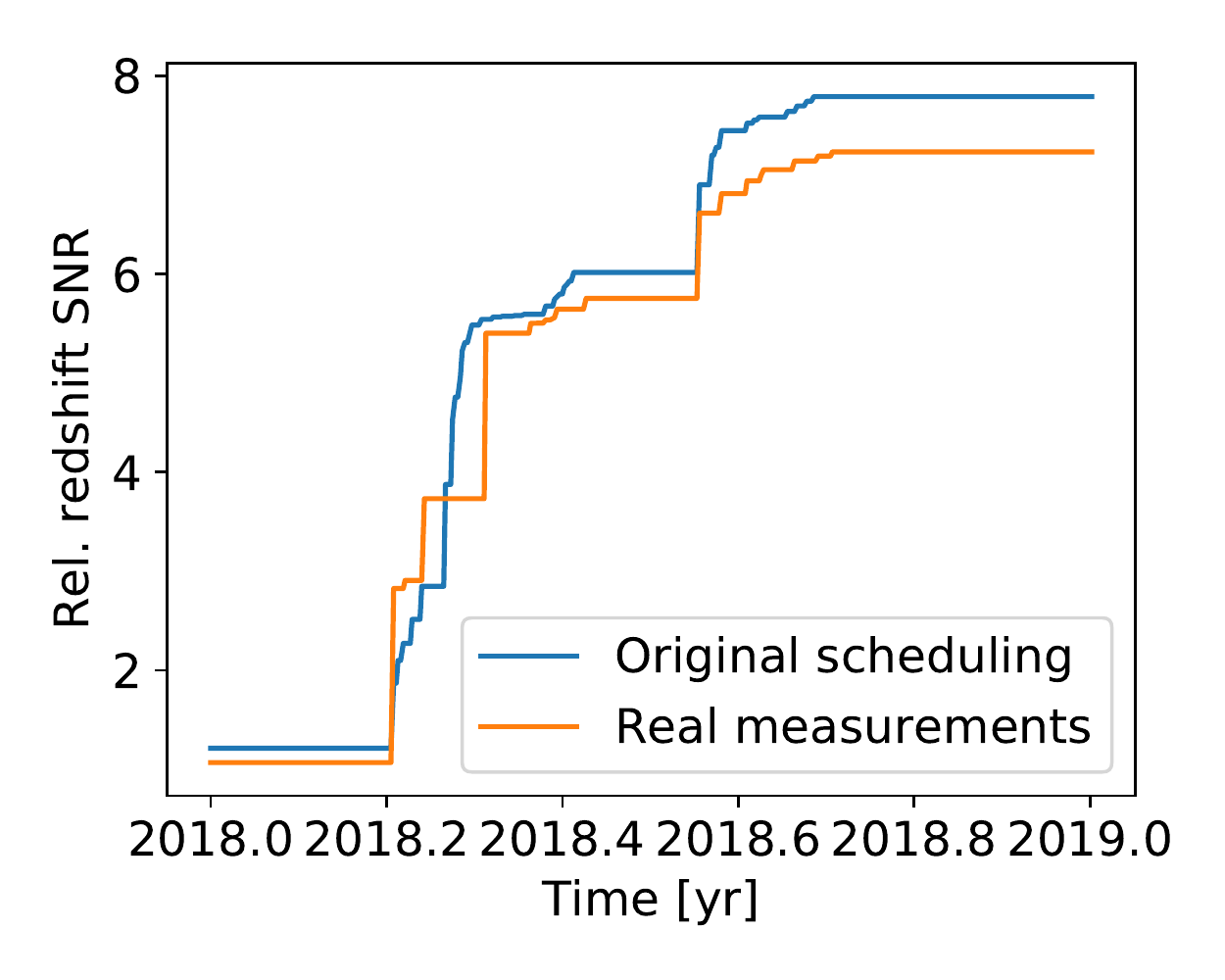}
\caption{\label{fig:red_SNR} Left: distribution of the relativistic redshift SNR after 2018 using the scheduling described in this section and assuming that each measurement has a 50\% probability to be lost. This SNR corresponds to the statistical uncertainty of the relativistic redshift, no systematics are considered here. Right: Evolution of the relativistic redshift SNR as a function of time. The blue curve represents the case corresponding to our original scheduling assuming all the measurements have been taken. The orange curve represents the evolution of the SNR obtained with the real measurements from 2018. These SNR corresponds to the statistical uncertainty of the redshift parameter, no systematics are considered in this analysis. Systematics have been thoroughly analyzed in \citet{do:2018aa}. }
\end{center}
\end{figure}

To summarize, the scheduling of the 2018 GC measurements campaign for the GCOI team was based on: (i) radial velocities measurements very early in 2018 to control our estimate of $T_0$ ; (ii) radial velocities at the two radial velocities turning points, the most important measurements to detect the redshift, using at least two different instruments to control for systematics ; (iii) measurements at closest approach important for other scientific objectives ; (iv) a dual spectroscopic-astrometric measurements at the important epochs to control for systematics ; (v) the total number of measurements has been determined by requiring that the relativistic redshift SNR at the end of 2018 has a probability of at least 99.9 \% to be above 5 considering that each night has a 50\% chance to be lost and (vi) the exact scheduling has been determined by considering a 3 days weather pattern. The scheduling resulting from this analysis is presented on the top of figure~\ref{fig:redshift_scheduling} and consists in 36 spectroscopic measurements (for Keck and Gemini) and 10 astrometric measurements. For the sake of comparison, the actual measurements performed in 2018 are presented in the lower panel from figure~\ref{fig:redshift_scheduling}. They consist in 19 radial velocity measurements and 5 astrometric measurements. The main sources of losses for 2018 are related to bad weather and to eruptions from the volcano in Hawaii, one of them being accompanied by an earthquake that damaged the instruments for a couple of days. The distribution of the relativistic redshift SNR at the end of 2018 using the scheduling described in this section and assuming that each measurement has a 50\% chance of being lost is presented on the right panel from figure~\ref{fig:red_SNR}. The probability to have a SNR below 5 is 0.03 \%. The SNR presented on this figure is a statistical SNR. No systematics is considered in this analysis and systematics will deter the SNR calculated here. The analysis of the real data has shown that including systematics such as offsets between instruments, correlations in the astrometric positional uncertainty or additional systematics uncertainties will produce a systematic uncertainty of the same order of magnitude as the statistical uncertainty (see \citet{do:2018aa} for the analysis of the measurements and a detailed study of systematics). The right panel of figure~\ref{fig:red_SNR} shows the evolution of the SNR with time in 2018. The blue curve corresponds to the evolution expected from the original scheduling while the orange curve corresponds to the evolution corresponding to the real measurements. Although nearly half of the measurements in 2018 have been lost, the final SNR obtained in the analysis remains above 5. This is due to the careful planning and to the number of measurements scheduled but also to the fact that some of the measurements were actually better than anticipated.

This first example shows the utility of the adaptive scheduling tool presented in section~\ref{sec:method} and how it can lead to surprising results. As a result of this analysis, the GCOI team decided to put more effort in measuring the radial velocity of S0-2 around the two turning points in order to maximize the chances of a successful detection of the relativistic redshfit~\citep{do:2018aa}. We also developed into details how the GCOI has decided to include systematics and possible weather/instrumental losses in the scheduling of the 2018 measurement campaign.

\subsection{Measurements on S0-2's orbit optimized to detect the relativistic advance of periastron}\label{sec:GR}
The relativistic redshift is only the first relativistic effect measurable using stellar orbits in the GC. The next effect expected to be measured is the well-known advance of periastron which is due to the first post-Newtonian relativistic correction to the equations of motion. In this section, we are interested in prioritizing measurements in order to measure this effect. For this reason, we fix the redshift parameter to its relativistic value (i.e. to unity) but we let the parameter that encodes a deviation from the first post-Newtonian equation of motion to be free. As a result, the model contains again 14 parameters. The set of ``existing measurements'' consists in astrometric and radial velocity measurements of S0-2 up to 2018. The set of future possible measurements consist in astrometric and spectroscopic measurements for each year between March and September (the yearly window for which GC measurements are feasible from Earth) on a daily basis. The uncertainty considered for the future measurements is of 0.2 mas for the astrometry and of 20 km/s for the radial velocity.

\begin{figure}
	\begin{center}
		\includegraphics[width=.9\hsize]{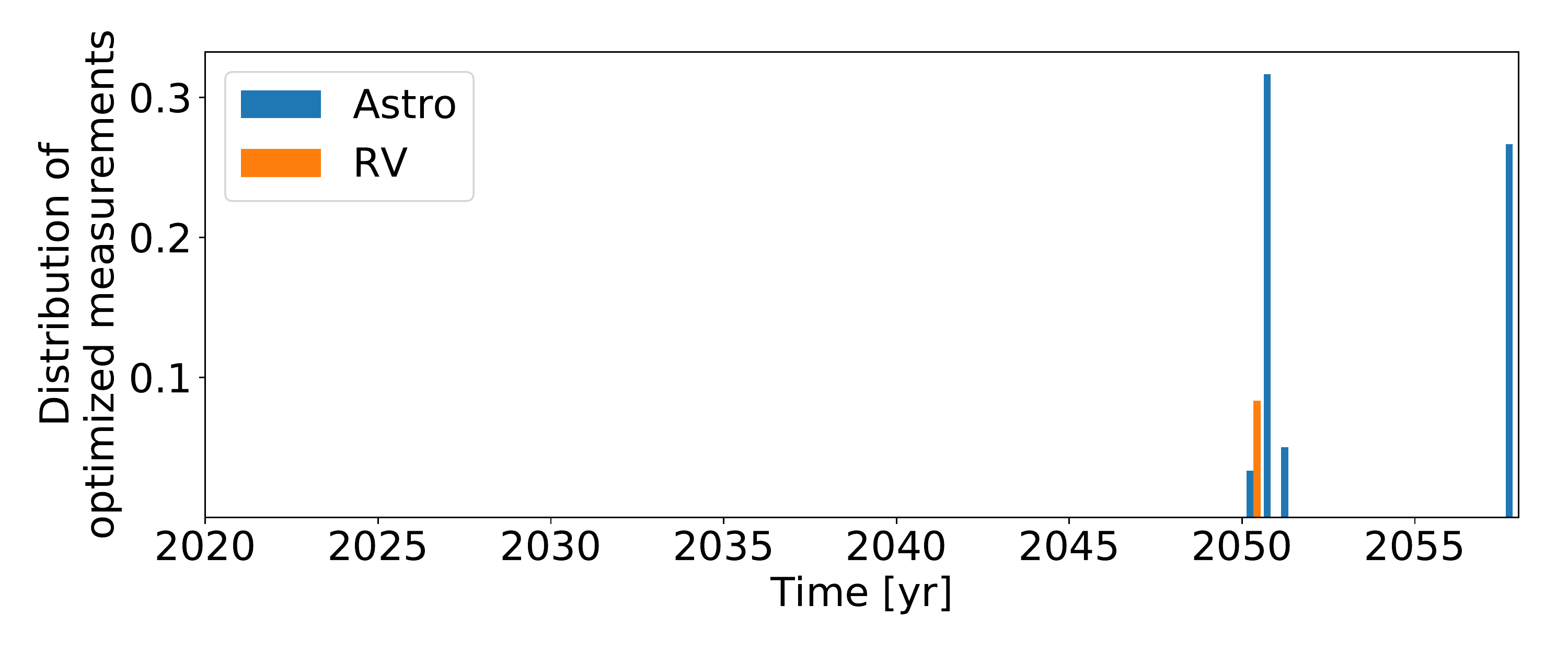}
		\caption{\label{fig:cadence_GR} Distribution of an optimized sequence of measurements in order to detect the relativistic advance of periastron on S0-2's orbit. This sequence of 30 measurements has been obtained considering possible observational windows that spans each year from March to September from 2018 to 2058. This figure shows three interesting features: (i) measuring the advance of the periastron relies mainly on astrometry measurements, (ii) late measurements are favored because it is a secular effect and (iii) measurements around closest approach are very sensitive to this effect.}
	\end{center}
\end{figure}

As a first step, we run the adaptive scheduling tool considering possible measurements between 2018 and 2058. The distribution the measurements from an optimized sequence of 30 observations is presented in figure~\ref{fig:cadence_GR}. While of limited interest, this figure shows three interesting features. First, it shows that, as anticipated, the detection of the advance of the periastron relies mainly on the astrometry. Secondly, it shows that the optimal strategy to measure the advance of the periastron given a certain measurement window is to take measurements as late as possible. This is easily explained considering that this effect is a secular effect that increases with time. Finally, we can notice that measurements around periastron (2050 correspond to a time of closest approach) are of prime importance. This last point will be developed further below. Changing the possible measurements window does not change qualitatively these three results. 

\begin{figure}
	\begin{center}
		\includegraphics[width=\hsize]{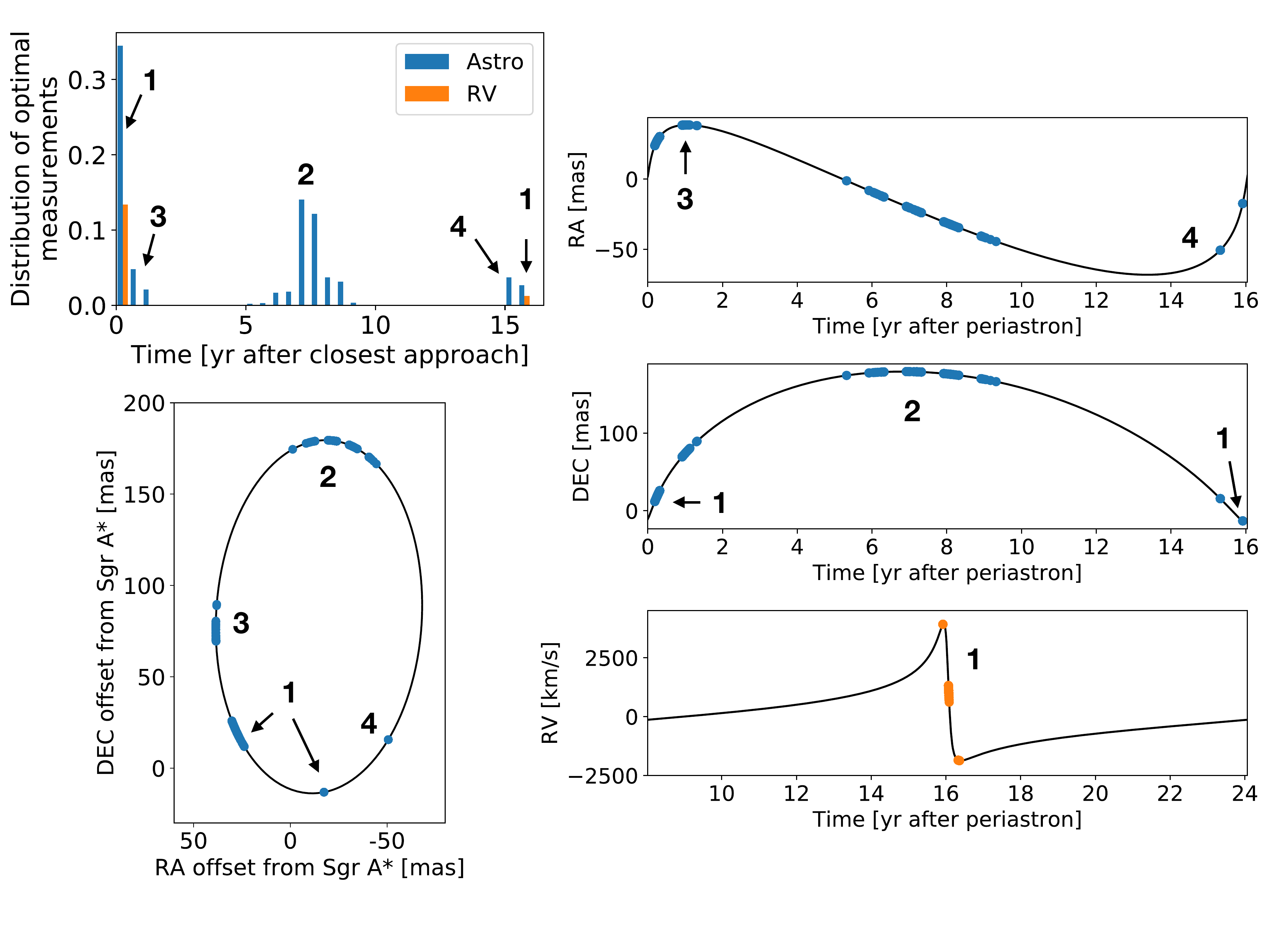}
		\caption{\label{fig:cadence_GR_2} Distribution of an optimized sequence of measurements in order to detect the relativistic advance of periastron on S0-2's orbit. Top left: distribution of the optimized measurements over one orbital period, each bin corresponding to half a year. The blue bins correspond to astrometric measurements while the orange bins correspond to the RV measurements. Bottom left: representation of the astrometric measurements (the same sequence as the top left panel) plotted in the plane of the sky. Right: representation of the measurements as a function of time: top: right ascension, middle: declination and bottom: RV. The numbers represent different set of astrometric measurements. 1: astrometric measurements bracketing closest approach, 2: astrometric measurements around apoastron, 3: astrometric measurements around the maximal RA turning point and 4:  astrometric measurements between the minimal right ascension turning point and closest approach.}
	\end{center}
\end{figure}

Figure~\ref{fig:cadence_GR} is somehow of limited interest because ``late'' measurements are always favored by the adaptive scheduling tool. A more interesting result consists in identifying the regions within one orbit that are particularly sensitive to the advance of the periastron. One way to derive such a result is to run the adaptive scheduling tool multiple times by increasing the measurements window one year at a time and to save all the optimized measurements sequence after removing the measurements that correspond to the last year of the observational window (to discard the ``late'' measurements effect due to the secular effect). The result from such an analysis is presented in figure~\ref{fig:cadence_GR_2} and is particularly interesting. The most sensitive measurements are astrometric measurements around closest approach. It is interesting to note that the closest approach itself is not favored by the adaptive scheduling tool (see the bottom left panel from figure~\ref{fig:cadence_GR_2}) because of correlations with other fitted parameters. Rather, it is important to have measurements bracketing the closest approach. The second important set of measurements consists of measurements around apoastron. This can easily be interpreted since this is the location where a precession of the orbit will have the largest impact. The next set of important measurements consists in the right ascension turning point (astrometric measurements around the maximum of the right ascension). In addition, while radial velocity measurement are less crucial for the detection of the relativistic advance of periastron, the important spectroscopic measurements are all located around the closest approach and at the two turning points in the radial velocity curve.

In order to quantify the impact of an optimized scheduling obtained using the algorithm from section~\ref{sec:method}, we compare the gain in the relativistic (advance of periastron) SNR after one orbital period for two different scheduling strategies. More precisely, we use the current existing measurements and compare two schedulings of 32 astrometric measurements for the next 16 years. The first scheduling consists in a ``naive'' sampling for which 2 measurements are taken every year and the second scheduling results from our optimization such that the measurements are timely distributed following the results presented in the top left panel from figure~\ref{fig:cadence_GR_2}. The SNR resulting from a regular sampling is of 4.1 while the SNR resulting from the optimized scheduling is of 5.9, a relative increase of 45\%. This shows that, although the scheduling resulting from the algorithm presented in section~\ref{sec:method} is not guaranteed to be the global optimum (see the discussion in section~\ref{sec:discussion}), it is nevertheless superior than standard non-optimized scheduling and it can be helpful for scientists to prioritize their measurements for a given scientific objective.

As a conclusion, the relativistic advance of the periastron is a secular effect such that long time baseline increases the chance for its detection. On top of that, measurements at the astrometric turning points (closest approach, apoastron and right ascension turning point) are important. In the mid-term, this means that 2019 will be an important year since it corresponds to the right ascension turning point and the years between 2022 and 2026 will be important to measure precisely the motion of S0-2 around the apoastron. 

\subsection{Measurements on S0-2's orbit optimized to improve our knowledge about the distance to the Galactic Center}\label{sec:R0}
In this section, we will focus on  the distance to our GC ($R_0s$), and prioritize  measurements to improve our knowledge of $R_0$ using the algorithm from section~\ref{sec:method}.  For this reason, we fix both GR parameters to their relativistic value (i.e. to unity) so that the model contains 13 free parameters. The set of ``existing measurements'' is exactly the same as the one used in the previous section, i.e. astrometric and radial velocity measurements of S0-2 up to 2018. The set of future possible measurements consist in astrometric and spectroscopic measurements for each year between March and September (the yearly window for which GC measurements are feasible from Earth) on a daily basis. The uncertainty considered for the future measurements is of 0.2 mas for the astrometry and of 20 km/s for the radial velocity.

\begin{figure}
	\begin{center}
		\includegraphics[width=\hsize]{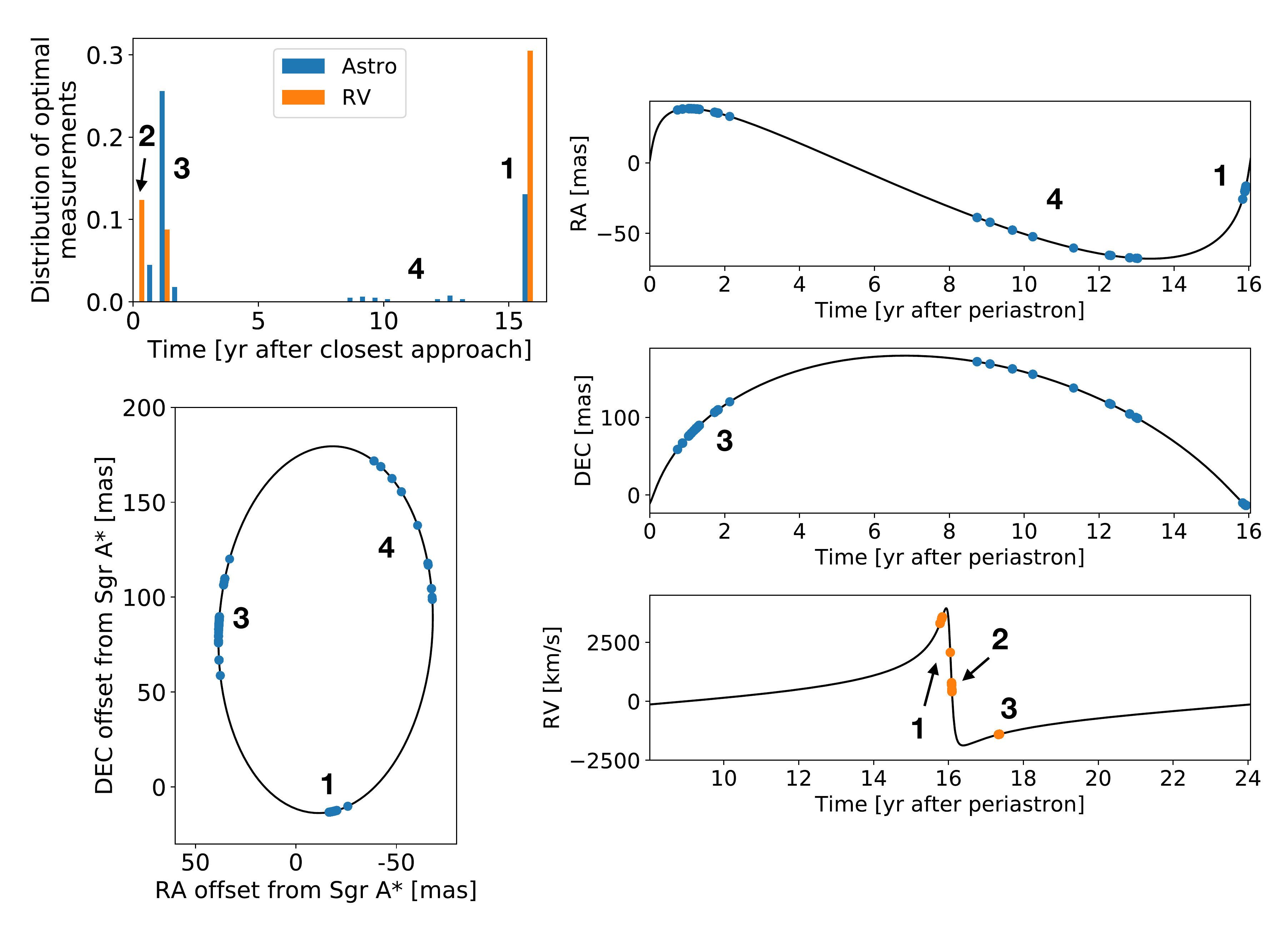}
		\caption{\label{fig:cadence_R0} Distribution of the measurements sequence optimized in order to improve our knowledge of the distance to the Galactic Center $R_0$ on S0-2's orbit.  Top left: distribution of the optimized  measurements over one orbital period, each bin corresponding to half a year. The blue bins correspond to astrometric measurements while the orange bins correspond to the RV measurements. Bottom left: representation the astrometric measurements plotted in the plane of the sky. Right: representation of the  measurements as a function of time: top: right ascension, middle: declination and bottom: RV. The numbers represent different set of  measurements. 1: astrometric and RV measurements just before closest approach, around RV max, 2: RV measurements around closest approach, 3: astrometric and RV measurements around the maximal right ascension turning point and 4: astrometric measurements after apoastron.}
	\end{center}
\end{figure}

First, a similar behavior as for the advance of the periastron is observed: late measurements are always favored by the adaptive scheduling tool and long time baseline will increase our knowledge on $R_0$. Therefore, we use the same strategy as the one described in the previous section: we run the adaptive scheduling tool by increasing the time baseline one year at a time, record the optimized measurements sequences but discard the observations related to the last year of the observational window. The result from such an analysis is presented in figure~\ref{fig:cadence_R0}. The estimation of $R_0$ relies on both astrometry and radial velocity. The important RV measurements are located at closest approach and are bracketing closest approach. The important astrometric measurements are located at the right ascension turning point (maximum of the right ascension curve) and before closest approach. Of particular interest, this analysis shows that the 2019 astrometric measurements will be of prime importance to improve our knowledge about $R_0$.

In order to assess the impact of the optimized scheduling, we compare the $R_0$ SNR between two schedulings extending over one orbital period: one regular sampling and one optimized scheduling using the algorithm from section~\ref{sec:method}. More precisely, we use the current existing measurements and compare two schedulings of 32 astrometric measurements and 16 RV for the next 16 years. The first scheduling consists in a ``naive'' scheduling for which 2 astrometric measurements and one radial velocity measurement are performed every year and the second scheduling results from our optimization such that the measurements are timely distributed following the results presented in the top left panel from figure~\ref{fig:cadence_R0}. The SNR resulting from a regular sampling is of 287 while the SNR resulting from the optimized scheduling is of 380, a relative increase of 30\%.  Similarly to what has been discussed in sections~\ref{sec:red} and \ref{sec:GR}, this shows the superiority of the scheduling obtained from the algorithm presented in section~\ref{sec:method} over standard non-optimized and uniform scheduling and it can help scientists to prioritize measurements for a given scientific objective.

\section{Conclusion}\label{sec:conclusion}
In this paper, we have presented a simple but efficient algorithm aiming at prioritizing measurements in order to reach a scientific objective. Identifying an optimized scheduling and the best type of measurements will become more and more important considering the increasing costs of measurements and considering the increasing pressure to obtain telescope time or time in different experimental facilities. The algorithm developed and presented in section~\ref{sec:method} relies on the assumption that the measurements errors and the posteriors are Gaussian but can be applied to any type of measurements or observations. This adaptive scheduling tool is based on the computation of the Fisher matrix, on a very efficient technique to update it and on a greedy algorithm. As discussed in section~\ref{sec:discussion}, a greedy algorithm is not guaranteed to converge to the global optimal solution but it has the advantage to have a reasonable time complexity, which makes it relatively fast and which allows the user to consider very large set of possible future measurements. In addition, the three examples discussed in section~\ref{sec:app} clearly shows the superiority of the optimized scheduling over non-optimized standard samplings (see also the discussion in \citet{loredo:2012aa}). 

We have used the case of measurements of the short-period star S0-2 in our GC as an example to illustrate the type of analysis and results that can be obtained with this adaptive scheduling tool. In particular, we have shown that the measurements that increase the the relativistic redshift SNR are radial velocity measurements at the turning points of the radial velocity curve and not at the maximum of the signal, a surprising result that is explained by correlations. We also suggest a method to plan measurements to control systematics and to account for possible weather/instrumental losses. This detailed analysis led to the scheduling for the GCOI 2018 measurement campaign detailed in section~\ref{sec:red}, which resulted in a successful detection of the relativistic redshift in 2018 (see \citet{do:2018aa}).

In addition, we have identified sets of measurements that are particularly important to detect the relativistic advance of periastron on S0-2's orbit. They consist of astrometric measurements at the astrometric turning points, in particular around periastron and apoastron. Finally, we have shown that both astrometry and radial velocity are important to improve our knowledge about $R_0$, the distance to our GC. The important astrometric measurements are the ones corresponding to the right ascension turning point and before closest approach while the  radial velocities adapted to measure $R_0$ are at closest approach and bracketing this epoch.

Similar analysis can be performed in various fields of physics and astrophysics ranging from Solar System measurements, exoplanets, gravitational waves and even laboratory measurements in order to help scientist to prioritize the type of measurements and their scheduling to increase their desired SNR in order to enhance the scientific return of different instruments and experiments.

\section{Acknowledgements}
Support for this work was provided by the the W.M. Keck Foundation (20170668), by the Heising Simons Foundation (2017-282) andby the National Science Foundation (AST-1412615). The authors warmly thank the anonymous referee whose comments helped to improve the manuscript.

\bibliographystyle{apj}
\bibliography{/Users/ahees/Documents/biblio/biblio}

\begin{thebibliography}{62}
\expandafter\ifx\csname natexlab\endcsname\relax\def\natexlab#1{#1}\fi

\bibitem[{{Albrecht} {et~al.}(2006){Albrecht}, {Bernstein}, {Cahn}, {Freedman},
  {Hewitt}, {Hu}, {Huth}, {Kamionkowski}, {Kolb}, {Knox}, {Mather}, {Staggs},
  \& {Suntzeff}}]{albrecht:2006aa}
{Albrecht}, A., {Bernstein}, G., {Cahn}, R., {Freedman}, W.~L., {Hewitt}, J.,
  {Hu}, W., {Huth}, J., {Kamionkowski}, M., {Kolb}, E.~W., {Knox}, L.,
  {Mather}, J.~C., {Staggs}, S., \& {Suntzeff}, N.~B. 2006, arXiv Astrophysics
  e-prints

\bibitem[{{Alexander}(2005)}]{alexander:2005ul}
{Alexander}, T. 2005, \physrep, 419, 65

\bibitem[{{Ang{\'e}lil} \& {Saha}(2010)}]{angelil:2010qd}
{Ang{\'e}lil}, R. \& {Saha}, P. 2010, \apj, 711, 157

\bibitem[{{Ang{\'e}lil} {et~al.}(2010){Ang{\'e}lil}, {Saha}, \&
  {Merritt}}]{angelil:2010to}
{Ang{\'e}lil}, R., {Saha}, P., \& {Merritt}, D. 2010, \apj, 720, 1303

\bibitem[{{Boehle} {et~al.}(2016){Boehle}, {Ghez}, {Sch{\"o}del}, {Meyer},
  {Yelda}, {Albers}, {Martinez}, {Becklin}, {Do}, {Lu}, {Matthews}, {Morris},
  {Sitarski}, \& {Witzel}}]{boehle:2016wu}
{Boehle}, A., {Ghez}, A.~M., {Sch{\"o}del}, R., {Meyer}, L., {Yelda}, S.,
  {Albers}, S., {Martinez}, G.~D., {Becklin}, E.~E., {Do}, T., {Lu}, J.~R.,
  {Matthews}, K., {Morris}, M.~R., {Sitarski}, B., \& {Witzel}, G. 2016, \apj,
  830, 17

\bibitem[{{Bourgoin} {et~al.}(2016){Bourgoin}, {Hees}, {Bouquillon}, {Le
  Poncin-Lafitte}, {Francou}, \& {Angonin}}]{bourgoin:2016yu}
{Bourgoin}, A., {Hees}, A., {Bouquillon}, S., {Le Poncin-Lafitte}, C.,
  {Francou}, G., \& {Angonin}, M.-C. 2016, Physical Review Letters, 117, 241301

\bibitem[{{Bourgoin} {et~al.}(2017){Bourgoin}, {Le Poncin-Lafitte}, {Hees},
  {Bouquillon}, {Francou}, \& {Angonin}}]{bourgoin:2017aa}
{Bourgoin}, A., {Le Poncin-Lafitte}, C., {Hees}, A., {Bouquillon}, S.,
  {Francou}, G., \& {Angonin}, M.-C. 2017, Physical Review Letters, 119, 201102

\bibitem[{{Do} {et~al.}(2019){Do}, {Hees.}, {Ghez}, {Martinez}, {Chu}, \& {et.
  al.}}]{do:2018aa}
{Do}, T., {Hees.}, A., {Ghez}, A.~M., {Martinez}, G., {Chu}, D., \& {et. al.}
  2019, {submitted to Science}

\bibitem[{{Eckart} \& {Genzel}(1997)}]{eckart:1997ys}
{Eckart}, A. \& {Genzel}, R. 1997, \mnras, 284, 576

\bibitem[{{Eckart} {et~al.}(2002){Eckart}, {Genzel}, {Ott}, \&
  {Sch{\"o}del}}]{eckart:2002qf}
{Eckart}, A., {Genzel}, R., {Ott}, T., \& {Sch{\"o}del}, R. 2002, \mnras, 331,
  917

\bibitem[{{Eisenhauer} {et~al.}(2005){Eisenhauer}, {Genzel}, {Alexander},
  {Abuter}, {Paumard}, {Ott}, {Gilbert}, {Gillessen}, {Horrobin}, {Trippe},
  {Bonnet}, {Dumas}, {Hubin}, {Kaufer}, {Kissler-Patig}, {Monnet},
  {Str{\"o}bele}, {Szeifert}, {Eckart}, {Sch{\"o}del}, \&
  {Zucker}}]{eisenhauer:2005dz}
{Eisenhauer}, F., {Genzel}, R., {Alexander}, T., {Abuter}, R., {Paumard}, T.,
  {Ott}, T., {Gilbert}, A., {Gillessen}, S., {Horrobin}, M., {Trippe}, S.,
  {Bonnet}, H., {Dumas}, C., {Hubin}, N., {Kaufer}, A., {Kissler-Patig}, M.,
  {Monnet}, G., {Str{\"o}bele}, S., {Szeifert}, T., {Eckart}, A.,
  {Sch{\"o}del}, R., \& {Zucker}, S. 2005, \apj, 628, 246

\bibitem[{{Eisenhauer} {et~al.}(2003){Eisenhauer}, {Sch{\"o}del}, {Genzel},
  {Ott}, {Tecza}, {Abuter}, {Eckart}, \& {Alexander}}]{eisenhauer:2003ty}
{Eisenhauer}, F., {Sch{\"o}del}, R., {Genzel}, R., {Ott}, T., {Tecza}, M.,
  {Abuter}, R., {Eckart}, A., \& {Alexander}, T. 2003, \apjl, 597, L121

\bibitem[{{Farnocchia} {et~al.}(2013){Farnocchia}, {Chesley}, {Chodas},
  {Micheli}, {Tholen}, {Milani}, {Elliott}, \& {Bernardi}}]{farnocchia:2013aa}
{Farnocchia}, D., {Chesley}, S.~R., {Chodas}, P.~W., {Micheli}, M., {Tholen},
  D.~J., {Milani}, A., {Elliott}, G.~T., \& {Bernardi}, F. 2013, \icarus, 224,
  192

\bibitem[{Fisher(1935)}]{fisher:1935aa}
Fisher, R.~A. 1935, Journal of the Royal Statistical Society, 98, 39

\bibitem[{{Ford}(2008)}]{ford:2008aa}
{Ford}, E.~B. 2008, \aj, 135, 1008

\bibitem[{{Gelman} {et~al.}(2013){Gelman}, {Carlin}, {Stern}, {Dunson},
  {Vehtari}, \& {Rubin}}]{gelman:2013aa}
{Gelman}, A., {Carlin}, J.~B., {Stern}, H.~S., {Dunson}, D.~B., {Vehtari}, A.,
  \& {Rubin}, D.~B. 2013, {Bayesian Data Analysis: 3rd Edition} ({Chapmann \&
  Hall/CRC press})

\bibitem[{{Genzel} {et~al.}(1997){Genzel}, {Eckart}, {Ott}, \&
  {Eisenhauer}}]{genzel:1997zr}
{Genzel}, R., {Eckart}, A., {Ott}, T., \& {Eisenhauer}, F. 1997, \mnras, 291,
  219

\bibitem[{{Ghez} {et~al.}(2003){Ghez}, {Duch{\^e}ne}, {Matthews}, {Hornstein},
  {Tanner}, {Larkin}, {Morris}, {Becklin}, {Salim}, {Kremenek}, {Thompson},
  {Soifer}, {Neugebauer}, \& {McLean}}]{ghez:2003qv}
{Ghez}, A.~M., {Duch{\^e}ne}, G., {Matthews}, K., {Hornstein}, S.~D., {Tanner},
  A., {Larkin}, J., {Morris}, M., {Becklin}, E.~E., {Salim}, S., {Kremenek},
  T., {Thompson}, D., {Soifer}, B.~T., {Neugebauer}, G., \& {McLean}, I. 2003,
  \apjl, 586, L127

\bibitem[{{Ghez} {et~al.}(2005{\natexlab{a}}){Ghez}, {Hornstein}, {Lu},
  {Bouchez}, {Le Mignant}, {van Dam}, {Wizinowich}, {Matthews}, {Morris},
  {Becklin}, {Campbell}, {Chin}, {Hartman}, {Johansson}, {Lafon}, {Stomski}, \&
  {Summers}}]{ghez:2005kx}
{Ghez}, A.~M., {Hornstein}, S.~D., {Lu}, J.~R., {Bouchez}, A., {Le Mignant},
  D., {van Dam}, M.~A., {Wizinowich}, P., {Matthews}, K., {Morris}, M.,
  {Becklin}, E.~E., {Campbell}, R.~D., {Chin}, J.~C.~Y., {Hartman}, S.~K.,
  {Johansson}, E.~M., {Lafon}, R.~E., {Stomski}, P.~J., \& {Summers}, D.~M.
  2005{\natexlab{a}}, \apj, 635, 1087

\bibitem[{{Ghez} {et~al.}(1998){Ghez}, {Klein}, {Morris}, \&
  {Becklin}}]{ghez:1998ve}
{Ghez}, A.~M., {Klein}, B.~L., {Morris}, M., \& {Becklin}, E.~E. 1998, \apj,
  509, 678

\bibitem[{{Ghez} {et~al.}(2000){Ghez}, {Morris}, {Becklin}, {Tanner}, \&
  {Kremenek}}]{ghez:2000rt}
{Ghez}, A.~M., {Morris}, M., {Becklin}, E.~E., {Tanner}, A., \& {Kremenek}, T.
  2000, \nat, 407, 349

\bibitem[{{Ghez} {et~al.}(2005{\natexlab{b}}){Ghez}, {Salim}, {Hornstein},
  {Tanner}, {Lu}, {Morris}, {Becklin}, \& {Duch{\^e}ne}}]{ghez:2005dq}
{Ghez}, A.~M., {Salim}, S., {Hornstein}, S.~D., {Tanner}, A., {Lu}, J.~R.,
  {Morris}, M., {Becklin}, E.~E., \& {Duch{\^e}ne}, G. 2005{\natexlab{b}},
  \apj, 620, 744

\bibitem[{{Ghez} {et~al.}(2008){Ghez}, {Salim}, {Weinberg}, {Lu}, {Do}, {Dunn},
  {Matthews}, {Morris}, {Yelda}, {Becklin}, {Kremenek}, {Milosavljevic}, \&
  {Naiman}}]{ghez:2008bs}
{Ghez}, A.~M., {Salim}, S., {Weinberg}, N.~N., {Lu}, J.~R., {Do}, T., {Dunn},
  J.~K., {Matthews}, K., {Morris}, M.~R., {Yelda}, S., {Becklin}, E.~E.,
  {Kremenek}, T., {Milosavljevic}, M., \& {Naiman}, J. 2008, \apj, 689, 1044

\bibitem[{{Gillessen} {et~al.}(2009){Gillessen}, {Eisenhauer}, {Fritz},
  {Bartko}, {Dodds-Eden}, {Pfuhl}, {Ott}, \& {Genzel}}]{gillessen:2009cr}
{Gillessen}, S., {Eisenhauer}, F., {Fritz}, T.~K., {Bartko}, H., {Dodds-Eden},
  K., {Pfuhl}, O., {Ott}, T., \& {Genzel}, R. 2009, \apjl, 707, L114

\bibitem[{{Gillessen} {et~al.}(2017){Gillessen}, {Plewa}, {Eisenhauer}, {Sari},
  {Waisberg}, {Habibi}, {Pfuhl}, {George}, {Dexter}, {von Fellenberg}, {Ott},
  \& {Genzel}}]{gillessen:2017aa}
{Gillessen}, S., {Plewa}, P.~M., {Eisenhauer}, F., {Sari}, R., {Waisberg}, I.,
  {Habibi}, M., {Pfuhl}, O., {George}, E., {Dexter}, J., {von Fellenberg}, S.,
  {Ott}, T., \& {Genzel}, R. 2017, \apj, 837, 30

\bibitem[{{Gravity Collaboration} {et~al.}(2018){Gravity Collaboration},
  {Abuter}, {Amorim}, {Anugu}, {Baub{\"o}ck}, {Benisty}, {Berger}, {Blind},
  {Bonnet}, {Brandner}, {Buron}, {Collin}, {Chapron}, {Cl{\'e}net}, {Coud{\'e}
  Du Foresto}, {de Zeeuw}, {Deen}, {Delplancke-Str{\"o}bele}, {Dembet},
  {Dexter}, {Duvert}, {Eckart}, {Eisenhauer}, {Finger}, {F{\"o}rster
  Schreiber}, {F{\'e}dou}, {Garcia}, {Garcia Lopez}, {Gao}, {Gendron},
  {Genzel}, {Gillessen}, {Gordo}, {Habibi}, {Haubois}, {Haug}, {Hau{\ss}mann},
  {Henning}, {Hippler}, {Horrobin}, {Hubert}, {Hubin}, {Jimenez Rosales},
  {Jochum}, {Jocou}, {Kaufer}, {Kellner}, {Kendrew}, {Kervella}, {Kok},
  {Kulas}, {Lacour}, {Lapeyr{\`e}re}, {Lazareff}, {Le Bouquin}, {L{\'e}na},
  {Lippa}, {Lenzen}, {M{\'e}rand}, {M{\"u}ler}, {Neumann}, {Ott}, {Palanca},
  {Paumard}, {Pasquini}, {Perraut}, {Perrin}, {Pfuhl}, {Plewa}, {Rabien},
  {Ram{\'{\i}}rez}, {Ramos}, {Rau}, {Rodr{\'{\i}}guez-Coira}, {Rohloff},
  {Rousset}, {Sanchez-Bermudez}, {Scheithauer}, {Sch{\"o}ller}, {Schuler},
  {Spyromilio}, {Straub}, {Straubmeier}, {Sturm}, {Tacconi}, {Tristram},
  {Vincent}, {von Fellenberg}, {Wank}, {Waisberg}, {Widmann}, {Wieprecht},
  {Wiest}, {Wiezorrek}, {Woillez}, {Yazici}, {Ziegler}, \&
  {Zins}}]{gravity:2018aa}
{Gravity Collaboration}, {Abuter}, R., {Amorim}, A., {Anugu}, N.,
  {Baub{\"o}ck}, M., {Benisty}, M., {Berger}, J.~P., {Blind}, N., {Bonnet}, H.,
  {Brandner}, W., {Buron}, A., {Collin}, C., {Chapron}, F., {Cl{\'e}net}, Y.,
  {Coud{\'e} Du Foresto}, V., {de Zeeuw}, P.~T., {Deen}, C.,
  {Delplancke-Str{\"o}bele}, F., {Dembet}, R., {Dexter}, J., {Duvert}, G.,
  {Eckart}, A., {Eisenhauer}, F., {Finger}, G., {F{\"o}rster Schreiber}, N.~M.,
  {F{\'e}dou}, P., {Garcia}, P., {Garcia Lopez}, R., {Gao}, F., {Gendron}, E.,
  {Genzel}, R., {Gillessen}, S., {Gordo}, P., {Habibi}, M., {Haubois}, X.,
  {Haug}, M., {Hau{\ss}mann}, F., {Henning}, T., {Hippler}, S., {Horrobin}, M.,
  {Hubert}, Z., {Hubin}, N., {Jimenez Rosales}, A., {Jochum}, L., {Jocou}, K.,
  {Kaufer}, A., {Kellner}, S., {Kendrew}, S., {Kervella}, P., {Kok}, Y.,
  {Kulas}, M., {Lacour}, S., {Lapeyr{\`e}re}, V., {Lazareff}, B., {Le Bouquin},
  J.-B., {L{\'e}na}, P., {Lippa}, M., {Lenzen}, R., {M{\'e}rand}, A.,
  {M{\"u}ler}, E., {Neumann}, U., {Ott}, T., {Palanca}, L., {Paumard}, T.,
  {Pasquini}, L., {Perraut}, K., {Perrin}, G., {Pfuhl}, O., {Plewa}, P.~M.,
  {Rabien}, S., {Ram{\'{\i}}rez}, A., {Ramos}, J., {Rau}, C.,
  {Rodr{\'{\i}}guez-Coira}, G., {Rohloff}, R.-R., {Rousset}, G.,
  {Sanchez-Bermudez}, J., {Scheithauer}, S., {Sch{\"o}ller}, M., {Schuler}, N.,
  {Spyromilio}, J., {Straub}, O., {Straubmeier}, C., {Sturm}, E., {Tacconi},
  L.~J., {Tristram}, K.~R.~W., {Vincent}, F., {von Fellenberg}, S., {Wank}, I.,
  {Waisberg}, I., {Widmann}, F., {Wieprecht}, E., {Wiest}, M., {Wiezorrek}, E.,
  {Woillez}, J., {Yazici}, S., {Ziegler}, D., \& {Zins}, G. 2018, \aap, 615,
  L15

\bibitem[{{Greenberg} {et~al.}(2017){Greenberg}, {Margot}, {Verma}, {Taylor},
  {Naidu}, {Brozovic}, \& {Benner}}]{greenberg:2017aa}
{Greenberg}, A.~H., {Margot}, J.-L., {Verma}, A.~K., {Taylor}, P.~A., {Naidu},
  S.~P., {Brozovic}, M., \& {Benner}, L.~A.~M. 2017, \aj, 153, 108

\bibitem[{{Gregory}(2010)}]{gregory:2010qv}
{Gregory}, P. 2010, {Bayesian Logical Data Analysis for the Physical Sciences}

\bibitem[{{Heavens}(2009)}]{heavens:2009aa}
{Heavens}, A. 2009, arXiv e-prints

\bibitem[{{Hees} {et~al.}(2017){Hees}, {Do}, {Ghez}, {Martinez}, {Naoz},
  {Becklin}, {Boehle}, {Chappell}, {Chu}, {Dehghanfar}, {Kosmo}, {Lu},
  {Matthews}, {Morris}, {Sakai}, {Sch{\"o}del}, \& {Witzel}}]{hees:2017aa}
{Hees}, A., {Do}, T., {Ghez}, A.~M., {Martinez}, G.~D., {Naoz}, S., {Becklin},
  E.~E., {Boehle}, A., {Chappell}, S., {Chu}, D., {Dehghanfar}, A., {Kosmo},
  K., {Lu}, J.~R., {Matthews}, K., {Morris}, M.~R., {Sakai}, S., {Sch{\"o}del},
  R., \& {Witzel}, G. 2017, Physical Review Letters, 118, 211101

\bibitem[{{Hees} {et~al.}(2015){Hees}, {Hestroffer}, {Le Poncin-Lafitte}, \&
  {David}}]{hees:2015rc}
{Hees}, A., {Hestroffer}, D., {Le Poncin-Lafitte}, C., \& {David}, P. 2015, in
  SF2A-2015: Proceedings of the Annual meeting of the French Society of
  Astronomy and Astrophysics, ed. F.~{Martins}, S.~{Boissier}, V.~{Buat},
  L.~{Cambr{\'e}sy}, \& P.~{Petit}, 125--131

\bibitem[{{Iorio}(2011)}]{iorio:2011rc}
{Iorio}, L. 2011, \mnras, 411, 453

\bibitem[{{Jaroszynski}(1998)}]{jaroszynski:1998xp}
{Jaroszynski}, M. 1998, \actaa, 48, 653

\bibitem[{Jaynes(2003)}]{jaynes:2003aa}
Jaynes, E.~T. 2003, Probability Theory: The Logic of Science, ed. G.~L.
  Bretthorst (Cambridge University Press)

\bibitem[{{Johannsen}(2016)}]{johannsen:2016sh}
{Johannsen}, T. 2016, Classical and Quantum Gravity, 33, 113001

\bibitem[{{Johannsen} {et~al.}(2016){Johannsen}, {Wang}, {Broderick},
  {Doeleman}, {Fish}, {Loeb}, \& {Psaltis}}]{johannsen:2016uq}
{Johannsen}, T., {Wang}, C., {Broderick}, A.~E., {Doeleman}, S.~S., {Fish},
  V.~L., {Loeb}, A., \& {Psaltis}, D. 2016, Physical Review Letters, 117,
  091101

\bibitem[{Kass \& Raftery(1995)}]{kass:1995aa}
Kass, R.~E. \& Raftery, A.~E. 1995, Journal of the American Statistical
  Association, 90, 773

\bibitem[{{Knuth} {et~al.}(2014){Knuth}, {Habeck}, {Malakar}, {Mubeen}, \&
  {Placek}}]{knuth:2014aa}
{Knuth}, K.~H., {Habeck}, M., {Malakar}, N.~K., {Mubeen}, A.~M., \& {Placek},
  B. 2014, arXiv e-prints

\bibitem[{{Konopacky} {et~al.}(2016){Konopacky}, {Marois}, {Macintosh},
  {Galicher}, {Barman}, {Metchev}, \& {Zuckerman}}]{konopacky:2016aa}
{Konopacky}, Q.~M., {Marois}, C., {Macintosh}, B.~A., {Galicher}, R., {Barman},
  T.~S., {Metchev}, S.~A., \& {Zuckerman}, B. 2016, \aj, 152, 28

\bibitem[{{Lainey} {et~al.}(2004){Lainey}, {Duriez}, \&
  {Vienne}}]{lainey:2004uq}
{Lainey}, V., {Duriez}, L., \& {Vienne}, A. 2004, A\&A, 420, 1171

\bibitem[{{Liddle}(2007)}]{liddle:2007aa}
{Liddle}, A.~R. 2007, \mnras, 377, L74

\bibitem[{{Loredo} {et~al.}(2010){Loredo}, {Chernoff}, {Liu}, {Clyde}, \&
  {Berger}}]{loredo:2010aa}
{Loredo}, T., {Chernoff}, D., {Liu}, B., {Clyde}, M., \& {Berger}, J. 2010, in
  ADA 6 - Sixth Conference on Astronomical Data Analysis, 13

\bibitem[{{Loredo}(2004)}]{loredo:2004aa}
{Loredo}, T.~J. 2004, in American Institute of Physics Conference Series, Vol.
  707, Bayesian Inference and Maximum Entropy Methods in Science and
  Engineering, ed. G.~J. {Erickson} \& Y.~{Zhai}, 330--346

\bibitem[{{Loredo} {et~al.}(2012){Loredo}, {Berger}, {Chernoff}, {Clyde}, \&
  {Liu}}]{loredo:2012aa}
{Loredo}, T.~J., {Berger}, J.~O., {Chernoff}, D.~F., {Clyde}, M.~A., \& {Liu},
  B. 2012, Statistical Methodology, 9, 101 , special Issue on Astrostatistics +
  Special Issue on Spatial Statistics

\bibitem[{{Loredo} \& {Chernoff}(2003)}]{loredo:2003aa}
{Loredo}, T.~J. \& {Chernoff}, D.~F. 2003, {Bayesian adaptive exploration}, ed.
  E.~D. {Feigelson} \& G.~J. {Babu}, 57--70

\bibitem[{{Ly} {et~al.}(2017){Ly}, {Marsman}, {Verhagen}, {Grasman}, \&
  {Wagenmakers}}]{ly:2017aa}
{Ly}, A., {Marsman}, M., {Verhagen}, J., {Grasman}, R., \& {Wagenmakers}, E.-J.
  2017, ArXiv e-prints

\bibitem[{{Moyer}(2003)}]{moyer:2000uq}
{Moyer}, T.~D. 2003, {Formulation for Observed and Computed Values of Deep
  Space Network Data Types for Navigation}, Deep-space communications and
  navigation series (Wiley-Interscience)

\bibitem[{{Nordtvedt}(1998)}]{nordtvedt:1998aa}
{Nordtvedt}, K. 1998, Classical and Quantum Gravity, 15, 3363

\bibitem[{{Parsa} {et~al.}(2017){Parsa}, {Eckart}, {Shahzamanian}, {Karas},
  {Zaja{\v c}ek}, {Zensus}, \& {Straubmeier}}]{parsa:2017aa}
{Parsa}, M., {Eckart}, A., {Shahzamanian}, B., {Karas}, V., {Zaja{\v c}ek}, M.,
  {Zensus}, J.~A., \& {Straubmeier}, C. 2017, \apj, 845, 22

\bibitem[{{Pihan-Le Bars} {et~al.}(2017){Pihan-Le Bars}, {Guerlin}, {Lasseri},
  {Ebran}, {Bailey}, {Bize}, {Khan}, \& {Wolf}}]{pihan-le-bars:2017aa}
{Pihan-Le Bars}, H., {Guerlin}, C., {Lasseri}, R.-D., {Ebran}, J.-P., {Bailey},
  Q.~G., {Bize}, S., {Khan}, E., \& {Wolf}, P. 2017, \prd, 95, 075026

\bibitem[{{Psaltis}(2008)}]{psaltis:2008uq}
{Psaltis}, D. 2008, Living Reviews in Relativity, 11, 9

\bibitem[{{Rubilar} \& {Eckart}(2001)}]{rubilar:2001dq}
{Rubilar}, G.~F. \& {Eckart}, A. 2001, \aap, 374, 95

\bibitem[{{Safronova} {et~al.}(2018){Safronova}, {Budker}, {DeMille},
  {Kimball}, {Derevianko}, \& {Clark}}]{safronova:2018aa}
{Safronova}, M.~S., {Budker}, D., {DeMille}, D., {Kimball}, D.~F.~J.,
  {Derevianko}, A., \& {Clark}, C.~W. 2018, Reviews of Modern Physics, 90,
  025008

\bibitem[{{Sakai} {et~al.}(2019){Sakai}, {Lu}, {Ghez}, {Jia}, {Do}, {Witzel},
  {Gautam}, {Hees}, {Becklin}, {Matthews}, \& {Hosek}}]{sakai:2019aa}
{Sakai}, S., {Lu}, J.~R., {Ghez}, A., {Jia}, S., {Do}, T., {Witzel}, G.,
  {Gautam}, A.~K., {Hees}, A., {Becklin}, E., {Matthews}, K., \& {Hosek}, Jr.,
  M.~W. 2019, \apj, 873, 65

\bibitem[{{Sch{\"o}del} {et~al.}(2002){Sch{\"o}del}, {Ott}, {Genzel},
  {Hofmann}, {Lehnert}, {Eckart}, {Mouawad}, {Alexander}, {Reid}, {Lenzen},
  {Hartung}, {Lacombe}, {Rouan}, {Gendron}, {Rousset}, {Lagrange}, {Brandner},
  {Ageorges}, {Lidman}, {Moorwood}, {Spyromilio}, {Hubin}, \&
  {Menten}}]{schodel:2002bh}
{Sch{\"o}del}, R., {Ott}, T., {Genzel}, R., {Hofmann}, R., {Lehnert}, M.,
  {Eckart}, A., {Mouawad}, N., {Alexander}, T., {Reid}, M.~J., {Lenzen}, R.,
  {Hartung}, M., {Lacombe}, F., {Rouan}, D., {Gendron}, E., {Rousset}, G.,
  {Lagrange}, A.-M., {Brandner}, W., {Ageorges}, N., {Lidman}, C., {Moorwood},
  A.~F.~M., {Spyromilio}, J., {Hubin}, N., \& {Menten}, K.~M. 2002, \nat, 419,
  694

\bibitem[{Sebastiani \& Wynn(2000)}]{sebastiani:2000aa}
Sebastiani, P. \& Wynn, H.~P. 2000, Journal of the Royal Statistical Society.
  Series B (Statistical Methodology), 62, 145

\bibitem[{{Vallisneri}(2008)}]{vallisneri:2008aa}
{Vallisneri}, M. 2008, \prd, 77, 042001

\bibitem[{{Verma} {et~al.}(2017){Verma}, {Margot}, \&
  {Greenberg}}]{verma:2017aa}
{Verma}, A.~K., {Margot}, J.-L., \& {Greenberg}, A.~H. 2017, \apj, 845, 166

\bibitem[{{Viswanathan} {et~al.}(2018){Viswanathan}, {Fienga}, {Minazzoli},
  {Bernus}, {Laskar}, \& {Gastineau}}]{viswanathan:2018aa}
{Viswanathan}, V., {Fienga}, A., {Minazzoli}, O., {Bernus}, L., {Laskar}, J.,
  \& {Gastineau}, M. 2018, \mnras, 476, 1877

\bibitem[{{Weinberg} {et~al.}(2005){Weinberg}, {Milosavljevi{\'c}}, \&
  {Ghez}}]{weinberg:2005cr}
{Weinberg}, N.~N., {Milosavljevi{\'c}}, M., \& {Ghez}, A.~M. 2005, \apj, 622,
  878

\bibitem[{{Will}(2008)}]{will:2008fk}
{Will}, C.~M. 2008, \apjl, 674, L25

\bibitem[{{Zucker} {et~al.}(2006){Zucker}, {Alexander}, {Gillessen},
  {Eisenhauer}, \& {Genzel}}]{zucker:2006fk}
{Zucker}, S., {Alexander}, T., {Gillessen}, S., {Eisenhauer}, F., \& {Genzel},
  R. 2006, \apjl, 639, L21

\end{thebibliography}

\end{document}